\input harvmac
\input tables

\def\epsK{\varepsilon_K}
\def\ra{\rightarrow}
\def\epspK{\varepsilon^\prime_K}
\def\ra{\rightarrow}
\def\CP{{\rm CP}}
\def\O{{\cal O}}

\def\Re{{\cal R}e}
\def\Im{{\cal I}m}
\def\gsim{{~\raise.15em\hbox{$>$}\kern-.85em
          \lower.35em\hbox{$\sim$}~}}
\def\lsim{{~\raise.15em\hbox{$<$}\kern-.85em
          \lower.35em\hbox{$\sim$}~}}

\def\gev{{\rm GeV}}

\def\H{{\cal H}}

\def\dKM{\delta_{\rm KM}}

\def\epsK{\varepsilon_K}

\noblackbox
\baselineskip 14pt plus 2pt minus 2pt
\Title{\vbox{\baselineskip12pt
\hbox{hep-ph/9810520}
\hbox{WIS-98/29/Oct-DPP}
}}
{\vbox{
\centerline{Flavor Physics and CP Violation}
  }}
\centerline{Yosef Nir}
\medskip
\centerline{\it Department of Particle Physics}
\centerline{\it Weizmann Institute of Science, Rehovot 76100, Israel}
\centerline{ftnir@clever.weizmann.ac.il}
\bigskip

\baselineskip 18pt
\noindent
This is a written version of a series of three lectures aimed at 
graduate students in the field of experimental high energy physics. 
The emphasis is on physics that is relevant to $B$-factories.
The main topics covered are:
\item{(i)} The flavor sector of the Standard Model;
\item{(ii)} Determination of the mixing parameters;
\item{(iii)} CP violation in meson decays (a model independent description);
\item{(iv)} CP violation in the Standard Model;
\item{(v)} CP violation as a probe of new physics.

\bigskip
\centerline{\it Lectures given at the}
\centerline{\bf 1998 European School of High Energy Physics}
\centerline{\it University of St. Andrews, Scotland}
\centerline{\it August 23 - September 5 1998} 

\Date{10/98}

\newsec{Flavor Physics}
\subsec{What is Flavor Physics and Why is it Interesting?}
The Standard Model fermions appear in three generations.
{\it Flavor physics} describes interactions that distinguish
between the fermion generations.
  
The fermions experience two types of interactions:
gauge interactions, where two fermions couple to a gauge boson,
and Yukawa interactions, where two fermions couple to a scalar.
Within the Standard Model
\nref\SMG{S.L. Glashow, Nucl. Phys. 22 (1961) 579.}%
\nref\SMW{S. Weinberg, Phys. Rev. Lett. 19 (1967) 1264.}%
\nref\SMS{A. Salam, in {\it Proc. 8th Nobel Symp.} (Stockholm),
 ed. N. Swartholm (Almquist and Wiksells, Stockholm 1968).}%
\refs{\SMG-\SMS}, there are twelve gauge bosons, related
to the gauge symmetry
\eqn\SMgg{G_{\rm SM}=SU(3)_{\rm C}\times SU(2)_{\rm L}\times U(1)_{\rm Y},}
and a single Higgs scalar, related to the spontaneous symmetry breaking
\eqn\SSB{G_{\rm SM}\rightarrow SU(3)_{\rm C}\times U(1)_{\rm EM}.}
In the {\it interaction basis}, gauge interactions are diagonal
(and universal, namely described by a single gauge coupling for each
factor in $G_{\rm SM}$: $g_s$, $g$ and $g^\prime$).
By definition, the interaction eigenstates have no gauge couplings 
between fermions of different generations. The Yuakawa interactions are, 
however, quite complicated in the interaction basis. In particular,
there are Yukawa couplings that involve fermions of different 
generations and, consequently, the interaction eigenstates
do not have well-defined masses. {\it Flavor Physics} refers 
to the part of the Standard Model that depends on the Yukawa couplings.

In the {\it mass basis}, Yukawa interactions are diagonal
(though not universal). The mass eigenstates have, by definition,
well-defined masses. The gauge interactions related to
spontaneously broken symmetries can, however, be quite complicated in the
mass basis. In particular, the $SU(2)_{\rm L}$ gauge couplings
are not diagonal, that is they {\it mix} quarks of different
generations. {\it Flavor Physics} here refers to fermion masses
and mixings.

Why is flavor physics interesting?
 
(i) Flavor physics has not been well tested yet. For the
gauge interactions, experiments (particularly LEP and SLD) have provided
us with tests at or even below an accuracy level of one percent,
where radiative corrections become essential. In contrast, several
flavor parameters are only known to an accuracy level of $\O(30\%)$.
Many rare decay processes that are sensitive to the flavor parameters
have not been measured yet. In the near future, various experiments 
(particularly CLEO, BaBar and Belle) will substantially improve
the determination of the flavor parameters and will measure various
rare $B$ decays, thus providing much
more stringent tests of this sector of the Standard Model.

(ii) Most of the Standard Model flavor parameters are small and hierarchical.
The Standard Model does not provide any explanation of these features.
This is {\it the flavor puzzle}. It may be a hint for physics beyond
the Standard Model, where the smallness and hierarchy of the flavor 
parameters find a natural explanation. For example, horizontal symmetries
(that is, symmetries under which different generations transform
differently) that are broken by a small parameter give selection rules
for the Yukawa couplings.

(iii) Flavor changing neutral current processes (FCNC) depend
on the flavor parameters. For vanishing Yukawa couplings,
FCNC would be absent to all orders in the gauge couplings.
Consequently, within the Standard Model FCNC are suppressed by
small mixing angles and, in some cases, small quark masses.
Furthermore, within the Standard Model FCNC vanish at tree level.
Consequently, they are further suppressed by powers of the weak coupling. 
Many extensions of the Standard Model allow significant new contributions 
to these processes that modify the Standard Model predictions. Therefore,
the flavor sector is a very sensitive probe of New Physics.

(iv) CP violation is closely related to flavor physics. 
It is one of the least tested aspects of the Standard Model: 
Even though the Kobayashi-Maskawa phase
\ref\KoMa{M. Kobayashi and T. Maskawa, Prog. Theo. Phys. 49 (1973) 652.}\
can account for the CP violation that has been measured in $K$ decays
\ref\CCFT{J.H. Christenson, J.W. Cronin, V.L. Fitch and R. Turlay,
 Phys. Rev. Lett. 13 (1964) 138.},
the Standard Model picture of CP violation could still be completely 
wrong. Almost any extension of the Standard Model provides new sources 
of CP violation. The observed baryon asymmetry of the universe 
{\it requires} new sources of CP violation
\ref\Sakh{A.D. Sakharov, ZhETF Pis. Red. 5 (1967) 32;
 JETP Lett. 5 (1967) 24.}.
(The motivation to study CP violation is described in more detail
in section 3.)

\subsec{What are the Flavor Parameters?}
 
The Standard Model fermions appear in three generations.
Each generation is made of five different representations
of the Standard Model gauge group $G_{\rm SM}$ of eq. \SMgg:
\eqn\SMrep{Q_{Li}^I(3,2)_{+1/6},\ \ u_{Ri}^I(3,1)_{+2/3},\ \ 
d_{Ri}^I(3,1)_{-1/3},\ \ L_{Li}^I(1,2)_{-1/2},\ \ \ell_{Ri}^I(1,1)_{-1}.}
Our notations mean that, for example, the left-handed quarks, $Q^I_L$, 
are in a triplet (3) of the $SU(3)_{\rm C}$ group, a doublet (2) of 
$SU(2)_{\rm L}$ and carry hypercharge $Y=Q_{\rm EM}-T_3=+1/6$. 
The index $I$ denotes {\it interaction eigenstates}. The 
index $i=1,2,3$ is the {\it flavor} (or generation) index.  
(The above representations describe quarks and leptons and
include, therefore, left-handed and right-handed fields.
An alternative way to write down the various representations,
which is particularly useful for the supersymmetric extension of
the Standard Model, is to describe left-handed fields only. Now
the fields include also antiquarks and antileptons:
\eqn\MSSMrep{Q_i^I(3,2)_{+1/6},\ \ \bar u_i^I(\bar3,1)_{-2/3},\ \ 
\bar d_i^I(\bar3,1)_{+1/3},\ \ L_i^I(1,2)_{-1/2},\ \ 
\bar\ell_i^I(1,1)_{+1}.)}

{\it The Standard Model gauge interactions
do not distinguish between the different generations.}
Another way to state this is to say that the
gauge interactions are flavor-blind. The strength of the
gauge interactions depends on the gauge quantum numbers
given in \SMrep\ and not on the flavor index $i$.
Most important for our purposes, the interaction of 
the $SU(2)_L$ gauge bosons ($W_\mu^a$, $a=1,2,3$) with quarks is given by
\eqn\Wint{-{\cal L}_W=
{g\over2}{\overline {Q^I_{Li}}}\gamma^\mu\tau^a Q^I_{Li} W_\mu^a.}
The $4\times4$ matrix $\gamma^\mu$ operates in Lorentz space 
(it describes the combination of two spin-1/2 quark fields and 
one spin-1 gauge boson field into a Lorentz scalar) and the $2\times2$
matrix $\tau^a$ operates in the $SU(2)_{\rm L}$ space (it describes
the combination of the two quark doublets and the $W^a$-triplet into
an $SU(2)_{\rm L}$ singlet). The coupling ${\overline {Q^I_{Li}}}Q^I_{Li}$
can be equivalently written as ${\overline {Q^I_{Li}}}{\bf1}_{ij}Q^I_{Lj}$
where the $3\times3$ unit matrix ${\bf1}$ operates in flavor space and makes 
the universality of the gauge interactions manifest.

The Yukawa interactions have a complicated form in this basis:
\eqn\Hint{-{\cal L}_Y=
Y^d_{ij}{\overline {Q^I_{Li}}}\phi d^I_{Rj}
+Y^u_{ij}{\overline {Q^I_{Li}}}\tilde\phi u^I_{Rj}
+Y^\ell_{ij}{\overline {L^I_{Li}}}\phi\ell^I_{Rj},}
where $\phi(1,2)_{+1/2}$ is the Standard Model Higgs doublet,
and $\tilde\phi=i\sigma_2\phi^*$. The Yukawa matrices
$Y^d$, $Y^u$ and $Y^\ell$ are general (and, in particular, complex)
$3\times3$ matrices. Note that, in the absence of right-handed neutrinos,
$N_i(1,1)_0$, one cannot write (renormalizable) Yukawa interactions
for the neutrinos.

To transform to the mass basis, one has to take into account
spontaneous symmetry breaking \SSB. Within the Standard Model
this breaking is the result of a vacuum expectation value assumed
by the neutral component of the Higgs doublet, $\vev{\phi^0}={v\over
\sqrt2}$ with the electroweak breaking scale of order $v\approx246\ GeV$.
Upon the replacement $\Re(\phi^0)\rightarrow(v+H^0)/\sqrt2$, 
the Yukawa interactions \Hint\ give rise to mass terms:
\eqn\Hint{-{\cal L}_M=
(M_d)_{ij}{\overline {d^I_{Li}}} d^I_{Rj}
+(M_u)_{ij}{\overline {u^I_{Li}}} u^I_{Rj}
+(M_\ell)_{ij}{\overline {\ell^I_{Li}}}\ell^I_{Rj},}
where
\eqn\YtoM{M_f={v\over\sqrt2}Y^f,}
and we decomposed the $SU(2)_{\rm L}$ doublets into their components:
\eqn\doublets{Q^I_{Li}=\pmatrix{u^I_{Li}\cr d^I_{Li}\cr},\ \ \ 
L^I_{Li}=\pmatrix{\nu^I_{Li}\cr \ell^I_{Li}\cr}.}
Since neutrinos have no Yukawa interactions, they are massless.

The mass basis corresponds, by definition, to diagonal mass matrices.
We can always find unitary matrices $V_{fL}$ and $V_{fR}$ such that
\eqn\diagM{V_{fL}M_f V_{fR}^\dagger=M_f^{\rm diag},}
with $M_f^{\rm diag}$ diagonal and real. The mass eigenstates
are then identified as
\eqn\masses{\eqalign{
d_{Li}=(V_{dL})_{ij}d_{Lj}^I&,\ \ \ d_{Ri}=(V_{dR})_{ij}d_{Rj}^I,\cr
u_{Li}=(V_{uL})_{ij}u_{Lj}^I&,\ \ \ u_{Ri}=(V_{uR})_{ij}u_{Rj}^I,\cr
\ell_{Li}=(V_{\ell L})_{ij}\ell_{Lj}^I&,\ \ \ 
\ell_{Ri}=(V_{\ell R})_{ij}\ell_{Rj}^I,\cr
\nu_{Li}=(V_{\nu L})_{ij}\nu_{Lj}^I&.\cr}}
Note that, since the neutrinos are massless, $V_{\nu L}$ is arbitrary.

The charged current interactions (that is the interactions of the charged 
$SU(2)_{\rm L}$ gauge bosons $W^\pm_\mu={1\over\sqrt2}
(W^1_\mu\mp iW_\mu^2)$), which in the interaction basis are described 
by \Wint, have a complicated form in the mass basis:
\eqn\Wmas{-{\cal L}_{W^\pm}=
{g\over\sqrt2}{\overline {u_{Li}}}\gamma^\mu(V_{uL}V_{dL}^\dagger)_{ij}
d_{Lj} W_\mu^++{\rm h.c.}.}
The $3\times3$ unitary matrix,
\eqn\VCKM{V_{\rm CKM}=V_{uL}V_{dL}^\dagger,} 
is the CKM {\it mixing matrix} for quarks
\nref\Cabi{N. Cabibbo, Phys. Rev. Lett. 10 (1963) 531.}%
\refs{\Cabi,\KoMa}. It generally depends on nine parameters:
three real angles and six phases. 

The form of the matrix is not unique. Usually, the following two  
conventions are employed:

$(i)$ There is freedom in defining $V_{\rm CKM}$ in that
we can permute between the various generations. This
freedom is fixed by ordering the up quarks and the down quarks
by their masses, i.e. $m_{u_1}< m_{u_2}<m_{u_3}$ and 
$m_{d_1}<m_{d_2}<m_{d_3}$. (Usually, we call $(u_1,u_2,u_3)\rightarrow
(u,c,t)$ and $(d_1,d_2,d_3)\rightarrow(d,s,b)$.) It is an interesting
fact that with this convention $V_{\rm CKM}$ is close to a unit matrix.
(See, for example, ref.
\ref\GrNiL{Y. Grossman and Y. Nir, Nucl. Phys. B448 (1995) 30.}\
for a discussion of this point in the framework of horizontal symmetries.)

$(ii)$ There is further freedom in the phase structure
of $V_{\rm CKM}$. Let us define $P_f$ ($f=u,d,\ell$) to be
diagonal unitary (phase) matrices. Then, if instead of using
$V_{fL}$ and $V_{fR}$ for the rotation \masses\ to the mass basis
we use $\tilde V_{fL}$ and $\tilde V_{fR}$, defined by
$\tilde V_{fL}=P_f V_{fL}$ and $\tilde V_{fR}=P_f V_{fR}$,
we still maintain a legitimate mass basis since $M_f^{\rm diag}$ remains
unchanged by such transformations. However, $V_{\rm CKM}$ does change:
\eqn\eqphase{V_{\rm CKM}\rightarrow P_u V_{\rm CKM}P_d^*.} 
This freedom is fixed by
demanding that $V_{\rm CKM}$ will have the minimal number of
phases. In the three generation case $V_{\rm CKM}$ has a single
phase. (There are five phase differences between the elements of
$P_u$ and $P_d$ and, therefore, five of the six phases in the
CKM matrix can be removed.) This is the Kobayashi-Maskawa phase 
$\delta_{\rm KM}$ which is the single source of 
{\it CP violation} in the Standard Model \KoMa.

As a result of the fact that $V_{\rm CKM}$ is not diagonal,
the $W^\pm$ gauge bosons can couple to quark (mass eigenstates)
of different generations. Within the Standard Model, this is
the only source of {\it flavor changing} interactions.
In principle, there could be additional sources of flavor
mixing in the lepton sector and in $Z^0$ interactions.
We now explain why, within the Standard Model, this does not happen.

{\it Mixing in the lepton sector}: An analysis similar to the above
applies also to the left-handed leptons. 
The mixing matrix is $(V_{\nu L}V_{\ell L}^\dagger)$. However, 
we can use the arbitrariness of $V_{\nu L}$ (related to the 
masslessness of neutrinos) to choose $V_{\nu L}=V_{\ell L}$, 
and the mixing matrix becomes a unit matrix. We conclude that 
the masslessness of neutrinos (if true) implies that there is 
no mixing in the lepton sector. If neutrinos have masses then 
the leptonic charged current interactions will exhibit mixing 
and CP violation.

{\it Mixing in neutral current interactions}:
Defining $\tan\theta_W\equiv g^\prime/g$, the Standard Model gives
\eqn\ZWB{Z^\mu=\cos\theta_W W_3^\mu-\sin\theta_W B^\mu.}
($B$ is the gauge boson related to $U(1)_{\rm Y}$.)
Therefore, to study the interactions of the $Z$ boson, we need to know 
the $W_3$-interactions (given in \Wint) and the $B$ interactions:
\eqn\Bint{-{\cal L}_B=-g^\prime\left[{1\over6}
{\overline{Q^I_{Li}}}\gamma^\mu{\bf1}_{ij}Q^I_{Lj}+
{2\over3}{\overline{u^I_{Ri}}}\gamma^\mu{\bf1}_{ij}u^I_{Rj}-
{1\over3}{\overline{d^I_{Ri}}}\gamma^\mu{\bf1}_{ij}d^I_{Rj}\right]B_\mu.}
Let us examine, for example, the $Z$-interactions with $d_L$ in
the mass basis:
\eqn\Zmas{\eqalign{-{\cal L}_Z=&{g\over\cos\theta_W}\left(-{1\over2}
+{1\over3}\sin^2\theta_W\right){\overline{d_{Li}}}
\gamma^\mu (V_{dL}^\dagger V_{dL})_{ij}d_{Lj}Z_\mu\cr
=&{g\over\cos\theta_W}\left(-{1\over2}
+{1\over3}\sin^2\theta_W\right){\overline{d_{Li}}}
\gamma^\mu d_{Li}Z_\mu.\cr}}
We learn that the neutral current interactions remain universal
in the mass basis and there are no additional flavor parameters
in their description. This situation goes beyond the Standard Model
to all models where all left-handed quarks are in $SU(2)_{\rm L}$
doublets and all right-handed ones in singlets. The $Z$-boson does
have flavor changing couplings in models where this is not the case. 

How many flavor parameters are there in the Standard Model?
In the interaction basis, the flavor parameters come from the
three Yukawa matrices. Since each of these is a $3\times3$
complex matrix, there are 27 real and 27 imaginary parameters
in these matrices. Not all of them are, however, physical.
If we switch off the Yukawa matrices, there is a global
symmetry added to the Standard Model, 
\eqn\Gglob{G_{\rm global}(Y^f=0)=U(3)_Q\times U(3)_{\bar d}
\times U(3)_{\bar u}\times U(3)_L\times U(3)_{\bar\ell}.}
A unitary rotation of the three generations for each of the 
five representations in \SMrep\ would leave the Standard Model 
Lagrangian invariant. This means that the physics described by 
a given set of Yukawa matrices $(Y^d,Y^u,Y^\ell)$, and the physics 
described by another set,
\eqn\rotY{\tilde Y^d=V_Q^\dagger Y^d V_{\bar d},\ \ \ 
\tilde Y^u=V_Q^\dagger Y^u V_{\bar u},\ \ \
\tilde Y^\ell=V_L^\dagger Y^\ell V_{\bar\ell},}
where $V$ are all unitary matrices, is the same.
One can use this freedom to remove, at most, 15 real and 30 
imaginary parameters (the number of parameters in five $3\times3$ 
unitary matrices). However, the fact that the Standard Model with
the Yukawa matrices switched on has still a global symmetry of
\eqn\SMglobal{G_{\rm global}=U(1)_B\times U(1)_e\times U(1)_\mu
\times U(1)_\tau}
means that only 26 imaginary parameters can be removed.
We conclude that there are 13 flavor parameters: 12 real ones
and a single phase. 

Examining the mass basis one can easily identify the flavor parameters.
In the quark sector, we have six quark masses, three mixing angles
(the number of real parameters in $V_{\rm CKM}$) and the single phase
$\delta_{\rm KM}$ mentioned above. In the lepton sector, we have
the three charged lepton masses.

\newsec{The Mixing Parameters}
While the fermion masses are determined from kinematics of various
processes so that the values are model independent, the mixing
parameters can only be determined from weak interaction processes
and could be affected by new physics. There is an intensive experimental 
effort to measure the elements of the CKM matrix, 
\eqn\VCKMij{V_{\rm CKM}=\pmatrix{V_{ud}&V_{us}&V_{ub}\cr
V_{cd}&V_{cs}&V_{cb}\cr V_{td}&V_{ts}&V_{tb}\cr}.}
There are three ways to determine the CKM parameters:
\item{(i)} Direct measurements: Standard Model tree level processes;
\item{(ii)} Unitarity: relations among the CKM elements
following from $V_{\rm CKM}^\dagger V_{\rm CKM}={\bf1}$;
\item{(iii)} Indirect measurements: Standard Model loop processes.

Direct measurements are expected to hold almost model independently.
The reason is that viable extensions of the Standard Model
have built-in mechanisms to suppress flavor changing processes
in order that the strong constraints from FCNC are satisfied.
These mechanisms make the contributions to Standard Model
tree level processes highly suppressed. In most extensions of the
Standard Model, the new physics takes place at a scale $\Lambda_{\rm NP}$
that is much higher than the EW breaking scale and consequently
the contributions to decay amplitudes are suppressed by
$\O(m_Z^2/\Lambda_{\rm NP}^2)\ll1$. In the next subsection we briefly 
describe the determination of CKM elements by direct measurements.

Unitarity holds if the only quarks (namely, color triplets with
electric charges $+2/3$ or $-1/3$) are those of the three generations
of the Standard Model. In many extensions of the Standard Model,
{\it e.g.} the minimal supersymmetric extension of the Standard Model,
this is indeed the situation and unitarity is 
a valid way of determining CKM elements. 

If there are additional quarks, which could be either sequential 
(fourth generation) or non-sequential ({\it e.g.} vector-like down 
quarks $D(3,1)_{-1/3}+\bar D(\bar3,1)_{+1/3}$), and if these extra 
quarks mix with the observed quarks, then CKM unitarity is violated. 

Indirect measurements are very sensitive to new physics. 
Take, for example, the $B-\bar B$ mixing amplitude. Within
the Standard Model, the leading contribution comes from
an EW box diagram and is therefore of ${\cal O}(g^4)$ and depends
on small mixing angles, $|V_{tb}V_{td}|^2$. These suppression
factors do not necessarily persist in extensions of the Standard
Model. For example, in supersymmetric models there could be
contributions of ${\cal O}(g_s^4)$ (gluino-mediated) and the
mixing angles could be comparable to (or even larger than) 
the Standard Model ones. The validity of indirect measurements 
is then model dependent.

One can make however a generic statement about the relation between
violation of CKM unitarity and the validity of indirect measurements
\ref\NiSia{Y. Nir and D. Silverman, Nucl. Phys. B345 (1990) 301.}.
Let us consider again the measurement of $|V_{tb}V_{td}|$ from $\Delta m_B$.
In models with vector-like down quarks, there is a 
{\it tree level} ($Z$-mediated)  contribution to this amplitude.
In four generation models, the heavy mass of the $t^\prime$ quark
gives an enhancement factor. In either case, whenever there is a
non-negligible violation of the CKM unitarity, there will be 
a much more significant modification of the Standard Model predictions 
through large contributions to FCNC processes.

The most efficient way to investigate the mixing parameters is 
then the following: 
\item{(1)} Measure as many parameters as possible by
direct measurements. At present we have $|V_{ud}|$, $|V_{us}|$,
$|V_{ub}|$, $|V_{cd}|$, $|V_{cs}|$, $|V_{cb}|$ and $|V_{tb}|$.
\item{(2)} Test whether the directly measured elements are
consistent with unitarity. If there is consistency,
determine the `missing' parameters (or improve the determination 
of those measured with large errors) by using unitarity. At present
we do so for $|V_{td}|$, $|V_{ts}|$, $|V_{tb}|$ and $|V_{cs}|$.
If there is inconsistency, then most likely the quark sector 
extends beyond the three generations of the Standard Model. 
\item{(3)} Test the predictions for FCNC processes. 
If there is consistency, one can further improve the determination
of poorly known CKM parameters. This is the case at present for
$|V_{tb}V_{td}|$ (from $\Delta m_B$ and $\Delta m_{B_s}$)
and for $\delta_{\rm KM}$ (from $\varepsilon_K$). 
If there is inconsistency, then New Physics has been discovered.

\subsec{Direct Measurements}
Seven of the nine absolute values of the CKM entries are measured
directly, namely by tree level processes.
(All numbers below are taken from 
\ref\PDG
{C. Caso {\it et al.}, Particle Data Group,
 Eur. Phys. Soc. J. C3 (1998) 1.}.)
Nuclear beta decays give
\eqn\VudExp{
|V_{ud}|=0.9740\pm0.0010.}
Semileptonic kaon and hyperon decays give
\eqn\VusExp{
|V_{us}|=0.2196\pm0.0023.}
Neutrino and antineutrino production of charm off valence $d$ quarks give
\eqn\VcdExp{
|V_{cd}|=0.224\pm0.016.}
Semileptonic $D$ decays give
\eqn\VcsExp{
|V_{cs}|= 1.04\pm0.16.}
Semileptonic exclusive and inclusive $B$ decays give
\eqn\VcbExp{
|V_{cb}|=0.0395\pm0.0017.}
The endpoint spectrum in semileptonic $B$ decays gives
\eqn\VubExp{
|V_{ub}/V_{cb}|=0.08\pm0.02.}
The decay $t\ra b\ell^+\nu_\ell$ gives
\eqn\VtbExp{
{|V_{tb}|^2/(|V_{tb}|^2+|V_{ts}|^2+|V_{td}|^2)}=0.99\pm0.29.}

\subsec{Unitarity of the CKM Matrix}
The requirement of CKM unitarity is simply stated as
$V_{\rm CKM}^\dagger V_{\rm CKM}={\bf1}$. This leads to various 
relations among the matrix elements. The orthogonality between
any two columns will be very useful in our discussion:
\eqn\Unitds{
V_{ud}V_{us}^*+V_{cd}V_{cs}^*+V_{td}V_{ts}^*=0,}
\eqn\Unitsb{
V_{us}V_{ub}^*+V_{cs}V_{cb}^*+V_{ts}V_{tb}^*=0,}
\eqn\Unitdb{
V_{ud}V_{ub}^*+V_{cd}V_{cb}^*+V_{td}V_{tb}^*=0.}
Another class of unitarity constraints is given by 
$\sum_{i=1}^3|V_{ij}|^2=\sum_{j=1}^3|V_{ij}|^2=1$.
A particularly useful relation is
\eqn\Unitbb{|V_{ub}|^2+|V_{cb}|^2+|V_{tb}|^2=1.}

Using unitarity constraints, one can narrow down some of the ranges
determined from direct measurements
(most noticeably, that of $|V_{cs}|$) and put constraints on the
top mixings $|V_{ti}|$. For example, the relation \Unitbb\ 
and the very small measured values of $|V_{ub}|$ and $|V_{cb}|$
imply that, to an excellent approximation,
\eqn\UnitVtb{|V_{tb}|=1.}
The relation \Unitsb\ and the small measured value of 
$|V_{us}V_{ub}|$ imply that, to a good approximation,
\eqn\UnitVts{|V_{ts}|\approx|V_{cb}|.}
The relation \Unitdb, together with $|V_{ub}/V_{cb}|\leq0.10$
and $|V_{cd}/V_{ud}|=0.22$, gives
\eqn\UnitVtd{|V_{td}V_{tb}|\approx0.0085\pm0.0045.}
The full information on the absolute values of the CKM elements 
from both direct measurements and three generation unitarity 
is summarized by \PDG:
\eqn\AbsCKM{
|V|=\pmatrix{0.9745-0.9760&0.217-0.224&0.0018-0.0045\cr
0.217-0.224&0.9737-0.9753&0.036-0.046\cr
0.004-0.013&0.035-0.042&0.9991-0.9994\cr}.}

The unitarity of the CKM matrix is manifest using an explicit
parameterization. There are various useful ways to parameterize it,
but the standard choice \PDG\ is the following
\ref\ChKe{L.L. Chau and W.Y. Keung, Phys. Rev. Lett. 53 (1984) 1802.}:
\eqn\CKCKM{
V=\pmatrix{c_{12}c_{13}&s_{12}c_{13}&s_{13}e^{-i\delta}\cr
-s_{12}c_{23}-c_{12}s_{23}s_{13}e^{i\delta}&
c_{12}c_{23}-s_{12}s_{23}s_{13}e^{i\delta}& s_{23}c_{13} \cr
s_{12}s_{23}-c_{12}c_{23}s_{13}e^{i\delta}&
-c_{12}s_{23}-s_{12}c_{23}s_{13}e^{i\delta}& c_{23}c_{13} \cr},}
where $c_{ij}\equiv\cos\theta_{ij}$ and $s_{ij}\equiv\sin\theta_{ij}$.
A test of the CKM picture is then whether there is a range for the
{\it four} parameters $s_{12}$, $s_{23}$, $s_{13}$ and $\delta$
that is consistent with the {\it seven} direct measurements described
in the previous subsection. Indeed, the following ranges are consistent
with \AbsCKM:
\eqn\CKcon{s_{12}=0.2196\pm0.0023,\ \ \ s_{23}=0.0395\pm0.0017,\ \ \ 
s_{13}/s_{23}=0.08\pm0.02.}
(The phase $\delta$ is not constrained at present by direct measurements.)
Another useful parametrization is in terms
of the four Wolfenstein parameters $(\lambda,A,\rho,\eta)$ with
$\lambda=|V_{us}|=0.22$ playing the role of an expansion parameter
and $\eta$ representing the CP violating phase
\ref\WOL{L. Wolfenstein, Phys. Rev. Lett. 51 (1983) 1945.}:
\eqn\WCKM{
V=\pmatrix{1-{\lambda^2\over2}&\lambda&A\lambda^3(\rho-i\eta)\cr
-\lambda&1-{\lambda^2\over2}&A\lambda^2\cr
A\lambda^3(1-\rho-i\eta)&-A\lambda^2&1\cr}+\O(\lambda^4).}
Because of the smallness of $\lambda$ and the fact that for each element
the expansion parameter is actually $\lambda^2$, it is sufficient to
keep only the first few terms in this expansion.
The ranges in \AbsCKM\ can be translated into the following ranges
of the Wolfenstein parameters:
\eqn\CKcon{\lambda=0.2196\pm0.0023,\ \ \ A=0.819\pm0.035,\ \ \ 
(\rho^2+\eta^2)^{1/2}=0.36\pm0.09.}
The relation between the parameters of \CKCKM\ and \WCKM\
is given by
\eqn\WoCK{
s_{12}\equiv\lambda,\ \ \ s_{23}\equiv A\lambda^2,\ \ \
s_{13}e^{-i\delta}\equiv A\lambda^3(\rho-i\eta).}
This specifies the higher order terms in \WCKM.

\subsec{Neutral Meson Mixing}
The presently useful indirect measurements ($\Delta m_B$,
$\Delta m_{B_s}$ and $\varepsilon_K$) are all related to
neutral meson mixing. Before presenting the implications
of these measurements for the CKM parameters, we briefly
discuss then the physics and formalism of neutral meson mixing.
We refer specifically to the neutral $B$ meson system, but 
most of our discussion applies equally well to the neutral $K$,
$B_s$ and $D$ systems.

Our phase convention for the CP transformation law of
the neutral $B$ mesons is defined by
\eqn\phacon{\CP|{B^0}\rangle=\omega_B|{\bar B^0}\rangle,\ \ \ 
\CP|{\bar B^0}\rangle=\omega_B^*|{B^0}\rangle,\ \ \ (|\omega_B|=1).}
Physical observables do not depend on the phase factor $\omega_B$.
An arbitrary linear combination of the neutral $B$-meson 
flavor eigenstates,
\eqn\defab{a|{B^0}\rangle+b|{\bar B^0}\rangle,}
is governed by a time-dependent Schr\"odinger equation,
\eqn\Schro{i{d\over dt}\pmatrix{a\cr b\cr}=H\pmatrix{a\cr b\cr}
\equiv\left(M-{i\over2}\Gamma\right)\pmatrix{a\cr b\cr},}
for which $M$ and $\Gamma$ are $2\times2$ Hermitian matrices.

The off-diagonal terms in these matrices, $M_{12}$ and $\Gamma_{12}$,
are particularly important in the discussion of mixing and CP violation.
$M_{12}$ is the dispersive part of the transition amplitude from
$B^0$ to $\bar B^0$. In the Standard Model it arises only at
order $g^4$. In the language of quark diagrams, the leading
contribution is from box diagrams. At sufficiently high loop momentum,
$k\gg\Lambda_{\rm QCD}$, these diagrams are a very good approximation
to the Standard Model contribution to $M_{12}$. This, or any other
contribution from heavy intermediate states from new physics, is
the {\it short distance} contribution. For small loop momenta,
$k\lsim1\ GeV$, we do not expect quark hadron duality to hold.
The box diagram is a poor approximation to the contribution from
light intermediate states, namely to {\it long distance} contributions.
Fortunately, in the $B$ and $B_s$ systems, the long distance contributions 
are expected to be negligible. (This is not the case for $K$ and $D$ 
mesons. Consequently, it is difficult to extract useful information 
from the measurement of $\Delta m_K$ and from the bound on $\Delta m_D$.)
$\Gamma_{12}$ is the absorptive part of the transition amplitude.
Since the cut of a diagram always involves on-shell particles and
thus long distance physics, the cut of the quark box diagram is a poor
approximation to $\Gamma_{12}$. However, it does correctly give
the suppression from small electroweak parameters such as the weak
coupling. In other words, though the hadronic uncertainties are large
and could change the result by order $50\%$, the cut in the 
box diagram is expected to give a reasonable order of magnitude estimate 
of $\Gamma_{12}$. (For $\Gamma_{12}(B_s)$ it has been shown that local
quark-hadron duality holds exactly in the simultaneous limit of small
velocity and large number of colors. We thus expect an uncertainty of
$\O(1/N_C)\sim30\%$
\nref\Alek{R. Aleksan {\it et al.}, Phys. Lett. B316 (1993) 567.}%
\nref\Bene{M. Beneke {\it et al.}, hep-ph/9808385.}%
\refs{\Alek-\Bene}. For $\Gamma_{12}(B_d)$ the small velocity limit
is not as good an approximation but an uncertainty of order 50\% still
seems a reasonable estimate.) New physics is not expected to affect 
$\Gamma_{12}$ significantly because it usually takes place at a high 
energy scale and is relevant to the short distance part only.

The light $B_L$ and heavy $B_H$ mass eigenstates are given by
\eqn\defqp{|{B_{L,H}}\rangle=p|{B^0}\rangle\pm q|{\bar B^0}\rangle.}
The complex coefficients $q$ and $p$ obey the normalization condition
$|q|^2+|p|^2=1$. Note that $\arg(q/p^*)$ is just an overall common
phase for $|B_L\rangle$ and $|B_H\rangle$ and has no physical significance.
The mass difference and the width difference between the 
physical states are given by
\eqn\DelmG{\Delta m\equiv M_H-M_L,\ \ \ \Delta\Gamma\equiv
\Gamma_H-\Gamma_L.}
Solving the eigenvalue equation gives
\eqn\eveq{\eqalign{(\Delta m)^2-{1\over4}(\Delta\Gamma)^2\ =&\
(4|M_{12}|^2-|\Gamma_{12}|^2),\cr
\Delta m\Delta\Gamma\ =&\ 4\Re(M_{12}\Gamma_{12}^*),\cr}}
\eqn\solveqp{{q\over p}=-{2M_{12}^*-i\Gamma_{12}^*\over
\Delta m-{i\over2}\Delta\Gamma}=-{\Delta m-{i\over2}\Delta\Gamma
\over 2M_{12}-i\Gamma_{12}}.}
In the $B$ system, $|\Gamma_{12}|\ll|M_{12}|$ (see discussion below),
and then, to leading order in $|\Gamma_{12}/M_{12}|$, \eveq\ and 
\solveqp\ can be written as
\eqn\eveqB{\Delta m_B=2|M_{12}|,\ \ \ 
\Delta\Gamma_B=\ 2\Re(M_{12}\Gamma_{12}^*)/|M_{12}|,}
\eqn\solveqpB{{q\over p}=-{M_{12}^*\over|M_{12}|}.} 

\subsec{Indirect Measurements} 
The most useful CP conserving indirect measurement is that of $\Delta m_B$
\ref\JAle{J. Alexander, talk given at 29th Int. Conf. on HEP in
Vancouver, July 22-29, 1998.}:
\eqn\xdExp{
\Delta m_B=0.471\pm0.016\ {\rm ps}^{-1}.}
The Standard Model accounts for $\Delta m_B=2|M_{12}|$ by box diagrams with
intermediate top quarks
\ref\InLi{
T. Inami and C.S. Lim, Prog. Theo. Phys. 65 (1981) 297; (E) 65 (1982) 772.}:
\eqn\MotSM{\Delta m_{B}={G_F^2\over6\pi^2}\eta_B m_Bm_W^2(B_Bf_B^2)
S_0(x_t)|V_{tb}V_{td}^*|^2,}
where $G_F$ is the Fermi constant, $\eta_B$ is a QCD correction factor 
calculated in NLO
\ref\BJW{A.J. Buras, M. Jamin and P.H. Weisz, Nucl. Phys. B347 (1990) 491.}, 
$S_0(x_t)$ is a kinematic function calculated 
from the box graphs \InLi, and $x_t=\bar m_t^2/m_W^2$. We use
\ref\BuFl{A.J. Buras and R. Fleischer, hep-ph/9704376, to appear
in {\it Heavy Flavours II}, eds. A.J. Buras and M. Lindner (World
Scientific).}\ 
${\overline{m_t}}(m_t)=167\pm6\ GeV$, giving $S_0(x_t)\approx2.36$. 
$B-\bar B$ mixing is dominated by
short distance physics (an intermediate top), so that the main source
of theoretical uncertainty lies in the matrix element of the four
quark operator between the meson states. The value of the matrix
element is parameterized by $B_Bf_B^2$ and is estimated by {\it e.g.}
lattice calculations \PDG, $B_Bf_B^2=(1.4\pm0.1)(175\pm25\ MeV)^2$.
The constraint on the CKM parameters from \MotSM\ can be written as
\eqn\xdCKM{|V_{tb}^*V_{td}|=0.0086
\left[{\Delta m_B\over0.471\ {\rm ps}^{-1}}\right]^{1/2}
\left[{0.2GeV\over\sqrt{B_B}f_B}\right]
\left[{2.4\over S_0(x_t)}\right]^{1/2}
\left[{0.55\over\eta_B}\right]^{1/2},}
This constraint gives at present
\eqn\xdVtd{|V_{tb}^*V_{td}|=0.0084\pm0.0018,}
which is consistent with, and actually significantly improves  
the unitarity constraint \UnitVtd.

Another useful indirect measurement if that of $\Delta m_{B_s}$.
The expression for $\Delta m_{B_s}$ is very similar to \xdExp, except
for the CKM dependence and an $SU(3)$ breaking factor (that is
an approximate global symmetry of the strong
interactions that holds in the limit $m_u=m_d=m_s=0$):
\eqn\BsBd{{\Delta m_{B_d}\over\Delta m_{B_s}} =
{m_{B_d}\over m_{B_s}}{B_{B_d}f_{B_d}^2\over B_{B_s}f_{B_s}^2}\
|V_{td}/V_{ts}|^2.}
The uncertainty in the ratio between the matrix elements is
smaller than the uncertainty in each of them separately
\ref\Sach{C.T. Sachrajda, in the {\it proceedings of the 18th International 
Symposium on Lepton Photon Interactions}, (Hamburg, 1997), hep-ph/9711386.}:
\eqn\UNCsd{{B_{B_s}f_{B_s}^2\over B_{B_d}f_{B_d}^2}=1.30\pm0.18.}
At present, there is only a lower bound \JAle, 
$\Delta m_{B_s}\geq12.4\ {\rm ps}^{-1}$, leading to
\eqn\VtdVts{|V_{td}/V_{ts}|\leq0.24\ \Longrightarrow\ |V_{td}|\leq0.0096,}
which further improves the upper bound of \xdVtd.

The imaginary part of the $K-\bar K$ mixing amplitude corresponds
to the CP violating observable $\epsK$ discussed in the next chapter:
\eqn\epsKdef{\epsK={\exp(i\pi/4)\over\sqrt2}{\Im\ M_{12}\over\Delta m_K}.}
The off-diagonal mass matrix element $M_{12}$ is obtained from the 
$\Delta S=2$ effective Hamiltonian with contributions from both the 
$c$-quark and the $t$-quark in the EW loop, yielding
\eqn\MotK{\eqalign{M_{12}=&{G_F^2\over12\pi^2}f_K^2B_Km_Km_W^2\cr
\times&\left[(V_{cd}^*V_{cs})^2\eta_1S_0(x_c)
+(V_{td}^*V_{ts})^2\eta_2S_0(x_t)+2(V_{cd}^*V_{cs})(V_{td}^*V_{ts})
\eta_3S_0(x_c,x_t)\right],\cr}}
where $f_K$ is the kaon decay constant and
$\eta_i$ are QCD factors calculated in NLO
\nref\HeNi{S. Herrlich and U. Niereste, Phys. Rev. D52 (1995) 6505;
Nucl. Phys. B476 (1996) 27.}%
\refs{\BJW,\HeNi}. $\Im\ M_{12}$
is dominated by short distance physics (intermediate top quark),
so that the main source of theoretical uncertainty lies in the matrix
element \Sach, $B_K=0.6-1$. The resulting constraint on the CKM parameters
can be written (with the convention that $(V_{ud}^*V_{us})$ is real) as
\eqn\epsKCKM{\eqalign{
\epsK=&\exp(i\pi/4)C_{\epsK}B_K\Im(V_{td}^*V_{ts})\cr\times&
\{\Re(V_{cd}^*V_{cs})[\eta_1S_0(x_c)-\eta_3S_0(x_c,x_t)]
-\Re(V_{td}^*V_{ts})\eta_2S_0(x_t)\},\cr}}
where all well-known quantities have been combined in the numerical constant,
\eqn\CepsK{C_{\epsK}={G_F^2\over6\sqrt2\pi^2}{f_K^2m_Km_W^2\over\Delta m_K}
=3.78\times10^4.} 

In the future, we may get useful information about the CKM parameters
from the two rare kaon decays, $K^+\ra\pi^+\nu\bar\nu$ 
\ref\BuBuc{G. Buchalla and A.J. Buras, Nucl. Phys. B412 (1996) 106.}\ 
and $K_L\ra\pi^0\nu\bar\nu$
\ref\BuBun{G. Buchalla and A.J. Buras, Nucl. Phys. B398 (1993) 285;
Nucl. Phys. B400 (1993) 225.}, 
which are theoretically very clean. Both modes are dominated by short 
distance $Z$-penguins and box diagrams. The branching ratio for $K^+\ra
\pi^+\nu\bar\nu$ can be expressed in terms of $\rho$ and $\eta$ \BuFl:
\eqn\BrKppnnc{BR(K^+\ra\pi^+\nu\bar\nu)=8.33\times10^{-6}
|V_{cb}|^4[X(x_t)]^2\left[\eta^2+(\rho_0-\rho)^2\right],}
where
\eqn\defrhpz{\rho_0=1+{P_0(X)\over X(x_t)}{\lambda^4\over|V_{cb}|^2},}
and $X(x_t)$ and $P_0(X)$ represent the electroweak loop contributions
in NLO for the top quark and for the charm quark, respectively.
The main theoretical uncertainty is related to the strong
dependence of the charm contribution on the renormalization
scale and the QCD scale, $P_0(X)=0.40\pm0.06$. First evidence
for $K^+\ra\pi^+\nu\bar\nu$ was presented recently
\ref\Adler{S. Adler {\it et al.}, E787 Collaboration, 
 Phys. Rev. Lett. 79 (1997) 2204.}.
The large experimental error does not yet give a useful CKM constraint
and is consistent with the Standard Model prediction.

The $K_L\ra\pi^0\nu\bar\nu$ decay is CP violating and will be discussed
later in detail. The branching ratio can be expressed in terms of $\eta$ 
\BuFl:
\eqn\BrKppnnn{BR(K_L\ra\pi^0\nu\bar\nu)=3.29\times10^{-5}
|V_{cb}|^4[X(x_t)]^2\eta^2.}
The present experimental bound, $BR(K_L\ra\pi^0\nu\bar\nu)\leq
1.6\times10^{-6}$
\ref\KTeV{J. Adams {\it et al.}, KTeV Collaboration, hep-ex/9806007.}\
lies about five orders of magnitude above the Standard Model prediction
\ref\BurWar{A.J. Buras, hep-ph/9610461.}\
and about two orders of magnitude above the bound that can be deduced
using model independent isospin relations
\ref\GrNi{Y. Grossman and Y. Nir, Phys. Lett. B398 (1997) 163.}\
from the experimental upper bound on the charged mode. 

\subsec{The Unitarity Triangle}
Each of the three relations \Unitds-\Unitdb\ requires the sum of 
three complex quantities to vanish and so can be geometrically 
represented in the complex plane as a triangle. These are 
``the unitarity triangles", though the term ``unitarity triangle" 
is usually reserved for the relation \Unitdb\ only. It is a surprising
feature of the CKM matrix that all unitarity triangles are equal in area.
For any choice of $i,j,k,l=1,2,3$, one can define a quantity $J$
according to
\ref\Jarl{C. Jarlskog, Phys. Rev. Lett. 55 (1985) 1039.}
\eqn\defJ{
\Im[V_{ij}V_{kl}V_{il}^*V_{kj}^*]=J\sum_{m,n=1}^3\epsilon_{ikm}
\epsilon_{jln}.}
Then, the area of each unitarity triangle equals $|J|/2$ while
the sign of $J$ gives the direction of the complex vectors
around the triangles. As will be discussed below, CP is violated 
in the Standard Model only if $J\neq0$. The area of the triangles
is then related to the size of the Standard Model CP violation.
 
The rescaled unitarity triangle  is derived from \Unitdb\
by (a) choosing a phase convention such that $(V_{cd}V_{cb}^*)$
is real, and (b) dividing the lengths of all sides by $|V_{cd}V_{cb}^*|$.
Step (a) aligns one side of the triangle with the real axis, and 
step (b) makes the length of this side 1. The form of the triangle 
is unchanged. Two vertices of the rescaled unitarity triangle are 
thus fixed at (0,0) and (1,0). The coordinates of the remaining 
vertex correspond to the Wolfenstein parameters $(\rho,\eta)$ (see \WCKM). 

Depicting the rescaled unitarity triangle in the
$(\rho,\eta)$ plane, the lengths of the two complex sides are
\eqn\RbRt{
R_u\equiv\sqrt{\rho^2+\eta^2}={1\over\lambda}
\left|{V_{ub}\over V_{cb}}\right|,\ \ \
R_t\equiv\sqrt{(1-\rho)^2+\eta^2}={1\over\lambda}
\left|{V_{td}\over V_{cb}}\right|.}
The three angles of the  unitarity triangle are denoted
by $\alpha,\beta$ and $\gamma$
\ref\DDGN{
C.O. Dib, I. Dunietz, F.J. Gilman and Y. Nir, Phys. Rev. D41 (1990) 1522.}:
\eqn\abcangles{
\alpha\equiv\arg\left[-{V_{td}V_{tb}^*\over V_{ud}V_{ub}^*}\right],\ \ \
\beta\equiv\arg\left[-{V_{cd}V_{cb}^*\over V_{td}V_{tb}^*}\right],\ \ \
\gamma\equiv\arg\left[-{V_{ud}V_{ub}^*\over V_{cd}V_{cb}^*}\right].}
They are physical quantities and, we will soon see, can be
independently measured by CP asymmetries in $B$ decays.

The only large uncertainties in the present determination
of the CKM elements are in $|V_{ub}|$ and $|V_{td}|$.
However, the two are related through \Unitdb. Thus, the unitarity
triangle is a very convenient tool for presenting constraints 
on these poorly determined parameters.
In particular, $\Delta m_{B_d}$ and $\Delta m_{B_s}$ constrain
$R_t$, the semileptonic $b\rightarrow u$ rates constrain $R_u$,
and the $\epsK$ constraint can be written as
\eqn\epsre{\eta\{(1-\rho)\eta_2S_0(x_t)|V_{cb}|^2+\eta_3S_0(x_c,x_t)-
\eta_1S_0(x_c)\}|V_{cb}|^2B_K=1.24\times10^{-6}.}

Examining the Wolfenstein parametrization, we learn that $|J|=\O(\lambda^6)
\times\sin\delta$. More precisely, the ranges specified above for 
the mixing angles give the following 90\% CL range: 
\eqn\standardrange{
|J|=(2.7\pm0.7)\times10^{-5}\ \sin\delta.}
The measurement of $\epsK$ is consistent with this range provided that
\eqn\eKdKM{\sin\delta=\O(1).}
(The phase $\delta$ is defined in eq. \CKCKM\ and equals $\gamma$
of eq. \abcangles.) When all available information, including the 
$\epsK$ constraint, is taken into account, we find the following 
allowed ranges for the CKM parameters
\nref\NirLP{Y. Nir, in the {\it proceedings of the 18th International 
Symposium on Lepton Photon Interactions}, (Hamburg, 1997), hep-ph/9709301.}%
\nref\BBPB{The BaBar Physics Book, to appear.}%
\nref\GNPS{Y. Grossman, Y. Nir, S. Plaszczynski and M.-H. Schune,
 Nucl. Phys. B511 (1998) 69, hep-ph/9709288.}%
\refs{\NirLP-\GNPS}:
\eqn\recon{
-0.15\leq\rho\leq+0.35,\ \ \ 
+0.20\leq\eta\leq+0.45,}
\eqn\abccon{
0.4\leq\sin2\beta\leq0.8,\ \ 
-0.9\leq\sin2\alpha\leq1.0,\ \ 
0.23\leq\sin^2\gamma\leq1.0.}

\newsec{CP Violation in Meson Decays: A Model Independent Discussion}
\subsec{Introduction}
CP violation arises naturally in the three generation Standard Model.
The CP violation that has been measured in neutral $K$-meson decays
($\epsK$) is accommodated in the Standard Model in simple way \KoMa.
Yet, CP violation is one of the least tested aspects of the 
Standard Model. The value of the $\epsK$ parameter \CCFT\ 
as well as bounds 
on other CP violating parameters (most noticeably, the electric
dipole moments of the neutron, $d_N$, and of the electron, $d_e$)
can be accounted for in models where CP violation has features
that are very different from the Standard Model ones. 

It is unlikely that the Standard Model provides the complete description
of CP violation in nature. First, it is quite clear that there exists
New Physics beyond the Standard Model. Almost any extension of the Standard
Model has additional sources of CP violating effects. In addition 
there is a great puzzle in cosmology that relates to CP violation, 
and that is the baryon asymmetry of the universe \Sakh.
Theories that explain the observed asymmetry
must include new sources of CP violation
\ref\CKN{For a review see, A.G. Cohen, D.B. Kaplan and
A.E. Nelson,  Ann. Rev. Nucl. Part. Sci. 43 (1993) 27.}: 
the Standard Model cannot
generate a large enough matter-antimatter imbalance to produce the
baryon number to entropy ratio observed in the universe today
\nref\FS{G.R. Farrar and M.E. Shaposhnikov, Phys. Rev. D50 (1994) 774.}%
\nref\Gave{M.B. Gavela {\it et al.}, Nucl. Phys. B430 (1994) 382.}%
\nref\HuSa{P. Huet and E. Sather, Phys. Rev. D51 (1995) 379.}%
\refs{\FS-\HuSa}.

In the near future, significant new information on CP violation
will be provided by various experiments. The main source of information
will be measurements of CP violation in various $B$ decays, particularly
neutral $B$ decays into final CP eigenstates
\nref\BCP{A.B. Carter and A.I. Sanda, Phys. Rev. Lett. 45 (1980) 952;
 Phys. Rev. D23 (1981) 1567.}%
\nref\BiSa{I.I. Bigi and A.I. Sanda, Nucl. Phys. B193 (1981) 85; 
 Nucl. Phys. B281 (1987) 41.}%
\nref\DuRo{
I. Dunietz and J. Rosner, Phys. Rev. D34 (1986) 1404.}%
\refs{\BCP-\DuRo}. Another piece of
valuable information might come from a measurement of the 
$K_L\ra\pi^0\nu\bar\nu$ decay
\nref\Litt{L.S. Littenberg, Phys. Rev. D39 (1989) 3322.}%
\nref\BuBuB{G. Buchalla and A.J. Buras, Phys. Lett. B333 (1994) 476.}%
\refs{\Litt,\BuBuB,\BuBun,\GrNi}. For the first time, the
pattern of CP violation that is predicted by the Standard Model
will be tested. Basic questions such as whether CP is an approximate
symmetry in nature will be answered.

It could be that the scale where new CP violating sources appear
is too high above the Standard Model scale ({\it e.g.} the GUT scale)
to give any observable deviations from the Standard Model predictions.
In such a case, the outcome of the experiments will be a (frustratingly)
succesful test of the Standard Model and a significant improvement in
our knowledge of the CKM matrix.

\nref\DLN{C.O. Dib, D. London and Y. Nir,
 Int. J. Mod. Phys. A6 (1991) 1253.}%
\nref\NiQuR{Y. Nir and H.R. Quinn,
 Ann. Rev. Nucl. Part. Sci. 42 (1992) 211.}%
\nref\GrLoR{M. Gronau and D. London, Phys. Rev. D55 (1997) 2845, 
 hep-ph/9608430.}%
A much more interesting situation will arise
if the new sources of CP violation appear at a scale 
that is not too high above the electroweak scale. Then they might be
discovered in the forthcoming experiments. Once enough independent
observations of CP violating effects are made, we will find that there
is no single choice of CKM parameters that is consistent with
all measurements. There may even be enough information in the pattern
of the inconsistencies to tell us something about the nature of the
new physics contributions \refs{\NiSia,\DLN-\GrLoR}.

The aim of this and the next two chapters is to explain the theoretical 
tools with which we will analyze new information about CP violation.
In this chapter, we give a brief, model-independent discussion
of CP violating observables. In the next chapter,
we discuss CP violation in the Standard Model.
In the last chapter, we describe CP violation beyond the Standard
Model and, in particular, in Supersymmetric models. The latter enables
us to elucidate the uniqueness of the Standard Model description
of CP violation and how little it has been tested so far. It further
demonstrates how the information from CP violation can help us probe
in detail models of New Physics.

\subsec{Notations and Formalism}

To understand the experimental and theoretical aspects of CP violation
in meson decays, we first introduce some formalism. We continue
the discussion of eqs. \phacon-\eveq. Again, we specifically discuss the 
neutral $B$ meson system, but large parts of our analysis apply 
equally well to the other meson systems.

To discuss CP violation in mixing (see below), 
it is useful to write \solveqpB\ to first order in $|\Gamma_{12}/M_{12}|$:
\eqn\solveqpB{{q\over p}=-{M_{12}^*\over|M_{12}|}\left[1-{1\over2}
\Im\left({\Gamma_{12}\over M_{12}}\right)\right].} 

To discuss CP violation in decay (see below), we need to consider
decay amplitudes. The CP transformation law for a final state $f$ is
\eqn\phaconf{\CP|{f}\rangle=\omega_f|{\bar f}\rangle,\ \ \ 
\CP|{\bar f}\rangle=\omega_f^*|{f}\rangle,\ \ \ (|\omega_f|)=1.}
For a final CP eigenstate $f=f_{\CP}$, the phase factor
$\omega_f$ is replaced by $\eta_{f_{\CP}}=\pm1$, the CP eigenvalue
of the final state.
We define the decay amplitudes $A_f$ and $\bar A_f$ according to
\eqn\defAf{A_f=\vev{f|{\cal H}_d|B^0},\ \ \ 
\bar A_f=\vev{f|{\cal H}_d|\bar B^0},}
where ${\cal H}_d$ is the decay Hamiltonian.  

To discuss CP violation in the interference of decays with and
without mixing (see below), we introduce a complex quantity
$\lambda_f$ defined by
\eqn\deflam{\lambda_f\ =\ {q\over p}\ {\bar A_f\over A_f}.}

The effective Hamiltonian that is relevant to $M_{12}$ is of the form
\eqn\Hbtwo{H^{\Delta b=2}_{\rm eff}\propto 
e^{+2i\phi_B}\left[\bar d\gamma^\mu(1-\gamma_5)b\right]^2
+e^{-2i\phi_B}\left[\bar b\gamma^\mu(1-\gamma_5)d\right]^2,}
where $2\phi_B$ is a CP violating (weak) phase. (We use the
Standard Model $V-A$ amplitude, but the results can be generalized
to any Dirac structure.) For the $B$ system, where $|\Gamma_{12}|\ll
|M_{12}|$, this leads to
\eqn\qpforB{q/p=\omega_B e^{-2i\phi_B}.}
(We implicitly assumed that the vacuum insertion approximation
gives the correct sign for $M_{12}$. In general, there is a
sign($B_B$) factor on the right hand side of \qpforB\
\ref\GKN{Y. Grossman, B. Kayser and Y. Nir,
 Phys. Lett. B415 (1997) 90, hep-ph/9708398.}.)
The decay Hamiltonian is of the form
\eqn\Hdecay{H_d\propto 
e^{+i\phi_f}\left[\bar q\gamma^\mu(1-\gamma_5)d\right]
\left[\bar b\gamma_\mu(1-\gamma_5)q\right]
+e^{-i\phi_f}\left[\bar q\gamma^\mu(1-\gamma_5)b\right]
\left[\bar d\gamma_\mu(1-\gamma_5)q\right],}
where $\phi_f$ is the appropriate weak phase. (Again, for simplicity
we use a $V-A$ structure, but the results hold for any Dirac structure.)
Then
\eqn\AbarA{\bar A_{\bar f}/A_f=\omega_f\omega_B^* e^{-2i\phi_f}.}
Eqs. \qpforB\ and \AbarA\ together imply that for a final CP eigenstate,
\eqn\lamfCP{\lambda_{f_{\CP}}=\eta_{f_{\CP}}e^{-2i(\phi_B+\phi_f)}.}

\subsec{The Three Types of CP Violation in Meson Decays}

There are three different types of CP violation in meson decays:
\item{(i)} CP violation in mixing, which occurs when the two neutral mass
eigenstate admixtures cannot be chosen to be CP-eigenstates;
\item{(ii)} CP violation in decay, which occurs in both charged and neutral
decays, when the amplitude for a decay and its CP-conjugate process
have different magnitudes;
\item{(iii)} CP violation in the interference of decays with and 
 without mixing, which occurs in decays into final states that are common to
$B^0$ and $\bar B^0$. (It often occurs in combination with the other two
types but there are cases when, to an excellent
approximation, it is the only effect.)

{\bf (i) CP violation in mixing:}
\eqn\inmixin{|q/p|\neq1.}
This results from the mass eigenstates being different from the
CP eigenstates, and requires a relative phase between $M_{12}$
and $\Gamma_{12}$. For the neutral $B$ system, this effect could
be observed through the asymmetries in semileptonic decays:
\eqn\mixexa{a_{\rm SL}={\Gamma(\bar B^0_{\rm phys}(t)\ra\ell^+\nu X)-
\Gamma(B^0_{\rm phys}(t)\ra\ell^-\nu X)\over
\Gamma(\bar B^0_{\rm phys}(t)\ra\ell^+\nu X)+
\Gamma(B^0_{\rm phys}(t)\ra\ell^-\nu X)}.}
In terms of $q$ and $p$,
\eqn\mixter{a_{\rm SL}={1-|q/p|^4\over1+|q/p|^4}.}
CP violation in mixing has been observed in the neutral $K$ system
($\Re\ \epsK\neq0$). 

In the neutral $B$ system, the effect is 
expected to be small, $\lsim\O(10^{-2})$. The reason is that,
model independently, the effect cannot be larger than
$\O(\Delta\Gamma_B/\Delta m_B)$. The difference in width is produced
by decay channels common to $B^0$ and $\bar B^0$. The branching
ratios for such channels are at or below the level of $10^{-3}$.
Since various channels contribute with differing signs, one expects
that their sum does not exceed the individual level. Hence
$\Delta\Gamma_B/\Gamma_B=\O(10^{-2})$ is a rather safe and
model independent assumption. On the other hand, it is
experimentaly known that $\Delta m_B/\Gamma_B\approx0.7$.

To calculate the deviation of $|q/p|$ from a pure phase (see \solveqpB),
\eqn\Absqp{
1-\left|{q\over p}\right|={1\over2}\Im{\Gamma_{12}\over M_{12}},} 
one needs to calculate $M_{12}$ and $\Gamma_{12}$. This involves 
large hadronic uncertainties, in particular in the hadronization 
models for $\Gamma_{12}$. 

{\bf (ii) CP violation in decay:}
\eqn\indecay{|\bar A_{\bar f}/A_f|\neq1.}
This appears as a result of interference among various terms in the
decay amplitude, and will not occur unless at least two terms have
different weak phases and different strong phases. CP asymmetries in
charged $B$ decays, 
\eqn\decexa{a_{f}={
\Gamma(B^+\ra f^+)-\Gamma(B^-\ra f^-)\over
\Gamma(B^+\ra f^+)+\Gamma(B^-\ra f^-)},}
are purely an effect of CP violation in decay. 
In terms of the decay amplitudes,
\eqn\decter{a_{f^\pm}={1-|\bar A_{f^-}/A_{f^+}|^2\over
1+|\bar A_{f^-}/A_{f^+}|^2}.}
There is as yet no unambiguous experimental evidence for CP violation
in decays. A measurement of $\Re\ \epsK^\prime\neq0$
\nref\NAto{G.D. Barr {\it et al.}, NA31 collaboration,
 Phys. Lett. B317 (1993) 233.}%
\nref\Esto{L.K. Gibbons {\it et al.}, E731 collaboration, 
 Phys. Rev. Lett. 70 (1993) 1203.}%
\refs{\NAto,\Esto}\ would constitute such evidence.
It is also possible that the first unambiguous evidence for
such CP violation will come from $B$ decays, {\it e.g.} for
$f^+=\pi^+ K^0$. 

There are two types of phases that may appear in $A_f$ and
$\bar A_{\bar f}$. Complex parameters in any Lagrangian term that 
contributes to the
amplitude will appear in complex conjugate form in the CP-conjugate
amplitude. Thus their phases appear in $A_f$ and $\bar A_{\bar f}$ with
opposite signs. In the Standard Model these phases occur only in the CKM
matrix which is part of the electroweak sector of the theory, hence these
are often called ``weak phases''. The weak phase of any single term is
convention dependent. However the difference between the weak phases in
two different terms in $A_f$ is convention independent because the phase
rotations of the initial and final states are the same for every term.
A second type of phase can appear in scattering or decay amplitudes even
when the Lagrangian is real. Such phases do not violate CP, since they
appear in $A_f$ and $\bar A_{\bar f}$ with the same sign. Their origin is
the possible contribution from intermediate on-shell states in the
decay process, that is an absorptive part of an amplitude that has
contributions from coupled channels. Usually the dominant
rescattering is due to strong interactions and hence the designation
``strong phases'' for the phase shifts so induced. Again only the
relative strong phases of different terms in a scattering amplitude
have physical content, an overall phase rotation of the entire
amplitude has no physical consequences.

Thus it is useful to write each contribution to $A$ in three parts:
its magnitude $A_i$; its weak phase term $e^{i\phi_i}$; and its strong
phase term $e^{i\delta_i}$. Then, if several amplitudes contribute to
$B\ra f$, we have
\eqn\defAtoA{\left|{\bar A_{\bar f}\over A_f}\right|=\left|{
\sum_i A_i e^{i(\delta_i-\phi_i)}\over  \sum_i A_i e^{i(\delta_i+\phi_i)}}
\right|.}
The magnitude and strong phase of any amplitude involve long distance
strong interaction physics, and our ability to calculate these from
first principles is limited. Thus quantities that depend only on the weak
phases are much cleaner than those that require knowledge of the
relative magnitudes or strong phases of various amplitude contributions,
such as CP violation in decay.
There is however a large literature and considerable theoretical effort
that goes into the calculation of amplitudes and strong phases . In many
cases we can only relate experiment to Standard Model parameters through
such calculations. The techniques that are used are expected to be more
accurate for $B$ decays than for $K$ decays, because of the larger $B$
mass, but theoretical uncertainty remains significant. The
calculations generally contain two parts. First the operator product
expansion and QCD perturbation theory are used to write any underlying
quark process as a sum of local quark operators with well-determined
coefficients. Then the matrix elements of the operators between the
initial and final hadron states must be calculated. This is where
theory is weakest and the results are most model dependent. Ideally
lattice calculations should be able to provide accurate determinations
for the matrix elements, and in certain cases this is already true, but
much remains to be done.

{\bf (iii) CP violation in the interference between decays
 with and without mixing:}
\eqn\ininter{|\lambda_{f_{\CP}}|=1,\ \ \Im\ \lambda_{f_{\CP}}\neq0.}
Any $\lambda_{f_{\CP}}\neq\pm1$ is a manifestation of CP violation. 
The special case \ininter\ isolates the effects of interest since both 
CP violation in decay \indecay\ and in mixing \inmixin\ lead to 
$|\lambda_{f_{\CP}}|\neq1$. For the neutral $B$ system, this effect can 
be observed by comparing decays into final CP eigenstates of a 
time-evolving neutral $B$ state that begins at time zero as $B^0$ 
to those of the state that begins as $\bar B^0$:
\eqn\intexa{a_{f_{\CP}}={
\Gamma(\bar B^0_{\rm phys}(t)\ra f_{\CP})-
\Gamma(B^0_{\rm phys}(t)\ra f_{\CP})\over
\Gamma(\bar B^0_{\rm phys}(t)\ra f_{\CP})+
\Gamma(B^0_{\rm phys}(t)\ra f_{\CP})}.}
This time dependent asymmetry is given (for $|\lambda_{f_{\CP}}|=1$) by
\eqn\intters{a_{f_{\CP}}=-\Im\lambda_{f_{\CP}}\sin(\Delta m_B t).}

CP violation in the interference of decays with and without
mixing has been observed for the neutral $K$ system
($\Im\ \epsK\neq0$). It is expected to be an effect of $\O(1)$
in various $B$ decays.  For such cases, the contribution from CP violation
in mixing is clearly negligible. For decays that are dominated
by a single CP violating phase (for example, $B\ra\psi K_S$ and
$K_L\ra\pi^0\nu\bar\nu$), so that the contribution from CP violation in
decay is also negligible, $a_{f_{\rm CP}}$ is cleanly interpreted
in terms of purely electroweak parameters. Explicitly, 
$\Im\lambda_{f_{\CP}}$ gives the difference between the phase of 
the $B-\bar B$ mixing amplitude ($2\phi_B$) and twice the phase of the 
relevant decay amplitude ($2\phi_f$) (see  eq. \lamfCP): 
\eqn\intCKM{\Im\lambda_{f_{\CP}}=-\eta_{f_{\CP}}\sin[2(\phi_B+\phi_f)].}

\subsec{The $\epsK$ Parameter}
Historically, a different language from the one used by us has been
employed to describe CP violation in $K\ra\pi\pi$ and $K\ra\pi\ell\nu$
decays. In this section we `translate' the language of $\varepsilon_K$
and $\varepsilon_K^\prime$ to our notations. Doing so will make it
easy to understand which type of CP violation is related to each quantity.

The two CP violating quantities measured in neutral $K$ decays are
\eqn\defetaij{
\eta_{00}={\vev{\pi^0\pi^0|\H|K_L}\over\vev{\pi^0\pi^0|\H|K_S}},\ \ \ 
\eta_{+-}={\vev{\pi^+\pi^-|\H|K_L}\over\vev{\pi^+\pi^-|\H|K_S}}.}
Define
\eqn\epsamp{\eqalign{
A_{00}=\vev{\pi^0\pi^0|\H|K^0},&\ \ \ 
\bar A_{00}=\vev{\pi^0\pi^0|\H|\bar K^0},\cr
A_{+-}=\vev{\pi^+\pi^-|\H|K^0},&\ \ \ 
\bar A_{+-}=\vev{\pi^+\pi^-|\H|\bar K^0},\cr}}
\eqn\epslam{
\lambda_{00}=\left({q\over p}\right)_K{\bar A_{00}\over A_{00}},\ \ \ 
\lambda_{+-}=\left({q\over p}\right)_K{\bar A_{+-}\over A_{+-}}.}
Then
\eqn\etapqA{
\eta_{00}={1-\lambda_{00}\over1+\lambda_{00}},\ \ \ 
\eta_{+-}={1-\lambda_{+-}\over1+\lambda_{+-}}.}
The $\eta_{00}$ and $\eta_{+-}$ parameters get contributions from 
CP violation in mixing ($|(q/p)|_K\neq1$) and from the interference 
of decays with and without mixing ($\Im\lambda_{ij}\neq0$) 
at $\O(10^{-3})$ and from CP violation in decay 
($|\bar A_{ij}/A_{ij}|\neq1$) at $\O(10^{-6})$.

There are two isospin channels in $K\ra\pi\pi$ leading to final
$(2\pi)_{I=0}$ and $(2\pi)_{I=2}$ states:
\eqn\twoiso{\eqalign{
\langle\pi^0\pi^0|=&\sqrt{1\over3}\langle(\pi\pi)_{I=0}|-
\sqrt{2\over3}\langle(\pi\pi)_{I=2}|,\cr
\langle\pi^+\pi^-|=&\sqrt{2\over3}\langle(\pi\pi)_{I=0}|+
\sqrt{1\over3}\langle(\pi\pi)_{I=2}|.\cr}}
The fact that there are two strong phases allows for CP violation
in decay. The possible effects are, however, small (on top of the 
smallness of the relevant CP violating phases) because the final
$I=0$ state is dominant (this is the $\Delta I=1/2$ rule). Defining
\eqn\defAI{A_I=\vev{(\pi\pi)_I|\H|K^0},\ \ \ 
\bar A_I=\vev{(\pi\pi)_I|\H|\bar K^0},}
we have, experimentally,
\eqn\AtwoAzer{|A_2/A_0|\approx1/20.}
Instead of $\eta_{00}$ and $\eta_{+-}$ we may define two combinations,
$\epsK$ and $\epspK$, in such a way that the possible effects of
CP violation in decay (mixing) are isolated into $\epspK$ ($\epsK$).

The experimental definition of the $\epsK$ parameter is 
\eqn\defepsex{\epsK\equiv{1\over3}(\eta_{00}+2\eta_{+-}).}
To zeroth order in $A_2/A_0$, we have $\eta_{00}=\eta_{+-}=\epsK$.
However, the specific combination \defepsex\ is chosen in such a way that 
the following relation holds to {\it first} order in $A_2/A_0$:
\eqn\defepsth{\epsK={1-\lambda_0\over1+\lambda_0}.}
Since, by definition, only one strong channel contributes to $\lambda_0$,
there is indeed no CP violation in decay in \defepsth.
It is simple to show that $\Re\ \epsK\neq0$ is a manifestation
of CP violation in mixing while $\Im\ \epsK\neq0$ is a manifestation
of CP violation in the interference between decays with and without 
mixing. Since experimentally $\arg\epsK\approx\pi/4$, the two
contributions are comparable.

The experimental definition of the $\epspK$ parameter is 
\eqn\defepspex{\epspK\equiv{1\over3}(\eta_{+-}-\eta_{00}).}
The theoretical expression is
\eqn\defepspth{\epspK\approx{1\over6}(\lambda_{00}-\lambda_{+-}).}
Obviously, any type of CP violation which is independent of the
final state does not contribute to $\epspK$. Consequently,
there is no contribution from CP violation
in mixing to \defepspth. It is simple to show that $\Re\ \epspK\neq0$ 
is a manifestation of CP violation in decay while $\Im\ \epspK\neq0$ 
is a manifestation of CP violation in the interference between decays 
with and without mixing.

\subsec{CP violation in $K\rightarrow\pi\nu\bar\nu$}
CP violation in the rare $K\ra\pi\nu\bar\nu$ decays is very
interesting. It is very different from the CP violation
that has been observed in $K\ra\pi\pi$ decays which is small and 
involves theoretical uncertainties. Similar to the CP asymmetry
in $B\ra\psi K_S$, it is predicted to be large and can be cleanly 
interpreted. Furthermore, observation of the $K_L\ra\pi^0\nu\bar\nu$
decay at the rate predicted by the Standard Model will provide
evidence that CP violation cannot be attributed to mixing 
($\Delta F=2$) processes only, as in superweak models. 
 
Define the decay amplitudes
\eqn\defApnn{A_{\pi^0\nu\bar\nu}=\vev{\pi^0\nu\bar\nu|\H|K^0},\ \ \
\bar A_{\pi^0\nu\bar\nu}=\vev{\pi^0\nu\bar\nu|\H|\bar K^0},}
and the related $\lambda_{\pi\nu\bar\nu}$ quantity:
\eqn\deflpnn{\lambda_{\pi\nu\bar\nu}=\left({q\over p}\right)_K
{\bar A_{\pi^0\nu\bar\nu}\over A_{\pi^0\nu\bar\nu}}.} 
The decay amplitudes of $K_{L,S}$ into a final $\pi^0\nu\bar\nu$
state are then
\eqn\KLpnn{\vev{\pi^0\nu\bar\nu|\H|\bar K_{L,S}}=pA_{\pi^0\nu\bar\nu}
\mp q\bar A_{\pi^0\nu\bar\nu},}
and the ratio between the corresponding decay rates is
\eqn\KLKSpnn{{\Gamma(K_L\ra \pi^0\nu\bar\nu)\over 
\Gamma(K_S\ra \pi^0\nu\bar\nu)}=
{1+|\lambda_{\pi\nu\bar\nu}|^2-2\Re\lambda_{\pi\nu\bar\nu}\over
1+|\lambda_{\pi\nu\bar\nu}|^2+2\Re\lambda_{\pi\nu\bar\nu}}.}
We learn that the $K_L\ra \pi^0\nu\bar\nu$ decay rate vanishes in
the CP limit ($\lambda_{\pi\nu\bar\nu}=1$), as expected on general
grouns \Litt.

Since the effects of CP violation in decay and in mixing are expected
to be negligibly small, $\lambda_{\pi\nu\bar\nu}$ is, to an excellent
approximation, a pure phase. Defining $\theta_K$ to be the relative
phase between the $K-\bar K$ mixing amplitude and the $s\ra d\nu\bar\nu$
decay amplitude, namely $\lambda_{\pi\nu\bar\nu}=e^{2i\theta_K}$, 
we get from \KLKSpnn:
\eqn\KLKSpn{{\Gamma(K_L\ra \pi^0\nu\bar\nu)\over 
\Gamma(K_S\ra \pi^0\nu\bar\nu)}={1-\cos2\theta_K\over1+\cos2\theta_K}=
\tan^2\theta_K.}
Using the isospin relation $A(K^0\ra\pi^0\nu\bar\nu)/
A(K^+\ra\pi^+\nu\bar\nu)=1/\sqrt2$, we get
\eqn\defapnn{a_{\pi\nu\bar\nu}\equiv{\Gamma(K_L\ra \pi^0\nu\bar\nu)\over 
\Gamma(K^+\ra \pi^+\nu\bar\nu)}={1-\cos2\theta_K\over2}=\sin^2\theta_K.}
Note that $a_{\pi\nu\bar\nu}\leq1$, and consequently a measurement
of $\Gamma(K^+\ra \pi^+\nu\bar\nu)$ can be used to set a model independent
upper limit on $\Gamma(K_L\ra \pi^0\nu\bar\nu)$ \GrNi.

\subsec{CP Violation in $D\rightarrow K\pi$ Decays}
Within the Standard Model, $D-\bar D$ mixing is expected to be well
below the experimental bound. Furthermore, effects related to CP violation
in $D-\bar D$ mixing are expected to be negligibly small since
this mixing is described to a good approximation by physics
of the first two generations. An experimental observation of
$D-\bar D$ mixing close to the present bound (and, obviously, of
related CP violation) will then be evidence for New Physics.
We now give the formalism of the neutral $D$ system for the case
that the mixing is close to the present bounds. 

We define the neutral $D$ mass eigenstates:
\eqn\defqpD{|{D_{1,2}}\rangle=p|{D^0}\rangle\pm q|{\bar D^0}\rangle.}
We define the following four decay amplitudes:
\eqn\defAKpi{\eqalign{A_{K^+\pi^-}=\vev{K^+\pi^-|{\cal H}|D^0},&\ \ \ 
\bar A_{K^+\pi^-}=\vev{K^+\pi^-|{\cal H}|\bar D^0},\cr
A_{K^-\pi^+}=\vev{K^-\pi^+|{\cal H}|D^0},&\ \ \ 
\bar A_{K^-\pi^+}=\vev{K^-\pi^+|{\cal H}|\bar D^0}.\cr}}
We introduce the following two quantities:
\eqn\deflpm{
\lambda_{K^+\pi^-}\ =\ \left({q\over p}\right)_D\ 
{\bar A_{K^+\pi^-}\over A_{K^+\pi^-}},\ \ \ 
\lambda_{K^-\pi^+}\ =\ \left({q\over p}\right)_D\ 
{\bar A_{K^-\pi^+}\over A_{K^-\pi^+}}.}

%
The following approximations are all experimentally confirmed:
\eqn\Dexp{\Delta m_D\ll\Gamma_D;\ \ \ \Delta \Gamma_D\ll\Gamma_D;\ \ \ 
|\lambda_{K^+\pi^-}^{-1}|\ll1;\ \ \ |\lambda_{K^-\pi^+}|\ll1.}
We are interested in the case that $\Delta m_D$ is found to be close 
to the present bound. (As mentioned above, this could only happen with 
new physics beyond the Standard Model.) Then, we can make a second 
approximation:
\eqn\MDGamD{\Delta \Gamma_D\ll \Delta m_D.}
We further make the reasonable assumptions that
CP violation in decay is negligible:
\eqn\negdec{\left|{A_{K^+\pi^-}\over\bar A_{K^-\pi^+}}\right|=
\left|{\bar A_{K^+\pi^-}\over A_{K^-\pi^+}}\right|=1,}
and that CP violation in mixing is negligible:
\eqn\negmix{\left|\left({q\over p}\right)_D\right|=1.}
With \negdec\ and \negmix, we find
\eqn\lamrel{|\lambda_{K^+\pi^-}^{-1}|=|\lambda_{K^-\pi^+}|\equiv
|\lambda_{K\pi}|.}

For the observation of mixing, we are interested in the state 
$|D^0(t)\rangle$ that starts out as a pure $|D^0\rangle$ at $t=0$ 
and in the state $|\bar D^0(t)\rangle$ that starts out as a pure 
$|\bar D^0\rangle$. The result of the above discussion is the 
following form for the (time dependent)
ratio between the doubly Cabibbo suppressed and Cabibbo allowed rates:
\eqn\DSCratio{\eqalign{
{\Gamma[D^0(t)\ra K^+\pi^-]\over\Gamma[D^0(t)\ra K^-\pi^+]}
=&|\lambda_{K\pi}|^2+{(\Delta m_D)^2\over4}t^2
+\Im(\lambda_{K^+\pi^-}^{-1})\Delta m_Dt,\cr
{\Gamma[\bar D^0(t)\ra K^-\pi^+]\over\Gamma[\bar D^0(t)\ra K^+\pi^-]}
=&|\lambda_{K\pi}|^2+{(\Delta m_D)^2\over4}t^2
+\Im(\lambda_{K^-\pi^+})\Delta m_Dt.\cr}}
(These are approximate expressions that hold for time 
$t\lsim{1\over\Gamma_D}$.)

The linear term is potentially CP violating. There are four possibilities
concerning this term
\nref\BSN{G. Blaylock, A. Seiden and Y. Nir,
 Phys. Lett. B355 (1995) 555, hep-ph/9504306.}%
\nref\WolD{L. Wolfenstein,
 Phys. Rev. Lett. 75 (1995) 2460, hep-ph/9505285.}%
\refs{\BSN,\WolD}:
\item{(i)} $\Im(\lambda_{K^+\pi^-}^{-1})=\Im(\lambda_{K^-\pi^+})=0$:
both strong and weak phases play no role in these processes.
\item{(ii)} $\Im(\lambda_{K^+\pi^-}^{-1})=\Im(\lambda_{K^-\pi^+})\neq0$:
weak phases play no role in these processes. There is a different
strong phase shift in $D^0\ra K^+\pi^-$ and $D^0\ra K^-\pi^+$.
\item{(iii)} $\Im(\lambda_{K^+\pi^-}^{-1})=-\Im(\lambda_{K^-\pi^+})\neq0$:
strong phases play no role in these processes. CP violating phases
affect the mixing amplitude.
\item{(iv)} $|\Im(\lambda_{K^+\pi^-}^{-1})|\neq|\Im(\lambda_{K^-\pi^+})|$:
both strong and weak phases play a role in these processes.

\newsec{CP Violation in The Standard Model}
\subsec{Introduction}
 
The irremovable phase in the CKM matrix allows CP violation.
Recalling the CP transformation laws,
\eqn\recallCPF{
\bar\psi_i\psi_j\ra\bar\psi_j\psi_i,\ \ \
\bar\psi_i\gamma^\mu W_\mu(1-\gamma_5)\psi_j\ra
\bar\psi_j\gamma^\mu W_\mu(1-\gamma_5)\psi_i,}
we learn that mass terms and gauge interactions can be CP invariant
if the masses and couplings are real. In particular, consider the
coupling of $W^\pm$ to quarks. It has the form
\eqn\Wquarks{
gV_{ij}\bar u_i\gamma_\mu W^{+\mu}(1-\gamma_5)d_j+
gV_{ij}^*\bar d_j\gamma_\mu W^{-\mu}(1-\gamma_5)u_i.}
The CP operation interchanges the two terms except that $V_{ij}$
and $V_{ij}^*$ are not interchanged. Thus, CP is a good symmetry
only if there is a mass basis where all couplings and masses are real.
 
CP is not necessarily violated in the three generation SM. If two
quarks of the same charge had equal masses, one mixing angle and the
phase could be removed from $V_{\rm CKM}$. Thus CP violation requires
\eqn\Jarlskogm{
(m_t^2-m_c^2)(m_c^2-m_u^2)(m_t^2-m_u^2)
(m_b^2-m_s^2)(m_s^2-m_d^2)(m_b^2-m_d^2)\neq0.}
If the value of any of the three mixing angles were 0 or $\pi/2$,
then again the phase could be removed. Finally, CP would not be
violated if the value of the single phase were 0 or $\pi$. These
last eight conditions are elegantly incorporated into one,
parameterization independent condition, that is (see \defJ\ for the
definition of $J$):
\eqn\JarlskogJ{J\neq0.}
(In the parameterization \CKCKM\
$J=c_{12}c_{23}c_{13}^2s_{12}s_{23}s_{13}\sin\delta$.
This shows explicitly that $J\neq0$ is equivalent to 
$\theta_{ij}\neq0,\pi/2$  and $\delta\neq0,\pi$.) 
The fourteen conditions incorporated in 
\Jarlskogm\ and \JarlskogJ\ can all be written as a single 
condition on the mass matrices in the interaction basis \Jarl:
\eqn\JarlCon{\Im\{\det[M_dM_d^\dagger,M_uM_u^\dagger]\}\neq0\ 
\Leftrightarrow\ \ {\rm CP\  violation}.}
The quantity $J$ is of much interest in the study of CP violation
from the CKM matrix. The maximum value that $J$ might assume is
$1/(6\sqrt3)\approx0.1$, but in reality it is $\sim3\times10^{-5}$.
  
Since the Standard Model contains only a single independent 
CP-violating phase, all possible CP-violating effects in this 
theory are very closely related. Consequently, the pattern of 
CP-violations in $B$ decays is strongly constrained. The goal 
of $B$ factories is to test whether this pattern occurs in Nature.
 
\subsec{CP Violation in Mixing}
In the $B_d$ system we expect that $\Gamma_{12}\ll M_{12}$ 
model independently.  Using the SM box diagrams to estimate 
the two quantities
\ref\BKUS{I.I. Bigi, V.A. Khoze, N.G. Uraltsev and A.I. Sanda, in {\it
$CP$ Violation}, ed. C. Jarlskog (World Scientific, Singapore, 1992).},
one gets 
\eqn\SMGtoM{
{\Gamma_{12}\over M_{12}}=-{3\pi\over2}{1\over f_2(m_t^2/m_W^2)}
{m_b^2\over m_t^2}\left(1+{8\over3}{m_c^2\over m_b^2}
{V_{cb}V_{cd}^*\over V_{tb}V_{td}^*}\right).}
This confirms our order of magnitude estimate, $|\Gamma_{12}/M_{12}|
\lsim10^{-2}$. CP violation in mixing is proportional
to $\Im(\Gamma_{12}/M_{12})$ which is even further suppressed:
\eqn\AbsqpSM{
1-\left|{q\over p}\right|={1\over2}\Im{\Gamma_{12}\over M_{12}}=
{4\pi\over f_2(m_t^2/m_W^2)}{m_c^2\over m_t^2}{J\over|V_{tb}V_{td}^*|^2}
\sim10^{-3}.}
Note that the suppression comes from the $(m_c^2/m_t^2)$ factor.
The last term is the ratio of the area of the unitarity triangle to
the length of one of its sides squared, so it is $\O(1)$. In contrast,
for the $B_s$ system, where \AbsqpSM\ holds except that $V_{td}$ is
replaced by $V_{ts}$, there is an additional suppression from
$J/|V_{tb}V_{ts}^*|^2\sim10^{-2}$ (see the corresponding 
unitarity triangle).

The above estimate of CP violation in mixing suffers from 
large uncertainties (of order 30\% \Alek\ or even higher
\ref\WolGot{L. Wolfenstein, Phys. Rev. D57 (1998) 5453.}) 
related to the use of a quark diagram to describe $\Gamma_{12}$ .

\subsec{CP Violation in Hadronic Decays of Neutral $B$}
In the previous subsection we estimated the effect of CP violation 
in mixing to be of $\O(10^{-3})$ within the Standard Model, and 
$\leq\O(|\Gamma_{12}/M_{12}|)\sim10^{-2}$ model independently
\ref\RaSu{For a recent discussion, see
 L. Randall and S. Su, hep-ph/9807377.}. 
In semileptonic decays, CP violation in mixing is the leading effect 
and therefore it can be measured through $a_{\rm SL}$. In purely hadronic 
$B$ decays, however, CP violation in decay and in the interference of 
decays with and without mixing is $\gsim \O(10^{-2})$. We can therefore 
safely neglect CP violation in mixing in the following discussion and use
\eqn\qpSM{{q\over p}=-{M_{12}^*\over|M_{12}|}=
{V_{tb}^*V_{td}\over V_{tb}V_{td}^*}\omega_B.} 

A crucial question is then whether CP violation
in decay is comparable to the CP violation in the interference of decays 
with and without mixing or negligible.  In the first case, we can
use the corresponding charged $B$ decays to observe effects of CP
violation in decay. In the latter case, CP asymmetries 
in neutral $B$ decays are subject to clean theoretical interpretation: 
we will either have precise measurements of CKM parameters or be provided 
with unambiguous evidence for new physics. The question of the relative
size of CP violation in decay can only be answered 
on a channel by channel basis, which is what we do in this section.

Most channels have contributions from both tree- and three types of
penguin-diagrams, the latter classified according to the
identity of the quark in the loop, as diagrams with different
intermediate quarks may have both different strong phases and different
weak phases
\ref\BSS{M. Bander, S. Silverman and A. Soni, 
 Phys. Rev. Lett. 43 (1979) 242.}. 
On the other hand, the subdivision of tree processes into
spectator, exchange and annihilation diagrams is unimportant in this
respect since they all carry the same weak phase.
 
While quark diagrams can be easily classified in this way, the
description of $B$ decays is not so neatly divided into tree and penguin
contributions once long distance physics effects are taken into
account. Rescattering processes can change the quark content of the final
state and confuse the identification of a contribution. There is no
physical distinction between rescattered tree diagrams and long-distance
contributions to the cuts of a penguin diagram.  While these issues
complicate estimates of various rates, they can always be avoided in
describing the weak phase structure of $B$-decay amplitudes. The decay
amplitudes for $b\rightarrow q \bar q q^\prime$ can always be
written as a sum of three terms with definite CKM coefficients:
\eqn\defCKM{
A(q \bar q q^\prime)=  V_{tb}V^*_{tq^\prime}P^t_{q^\prime} +
V_{cb}V_{cq^\prime}^*(T_{c\bar c q^\prime}\delta_{qc} + P^c_{q^\prime})+
V_{ub}V_{uq^\prime}^*(T_{u \bar u q^\prime}\delta_{qu} + P^u_{q^\prime}).}
Here $P $ and $T$ denote contributions from tree and penguin diagrams,
excluding the CKM factors. As they stand, the $P$ terms are not
well defined because of the divergences of the penguin diagrams.  Only
differences of penguin diagrams are finite and well defined. However
already we see that diagrams that can be mixed by rescattering effects
always appear with the same CKM coefficients and hence that a separation
of these terms is not needed when discussing weak phase structure.
Now it is useful to use eqs. \Unitsb\ and \Unitdb\
to eliminate one of the three
terms, by writing its CKM coefficient as minus the sum of the other two.
 
In the case of $q \bar q s$ decays it is convenient to remove the
$V_{tb} V^*_{ts}$ term. Then
\eqn\qqstype{\eqalign{
A(c\bar c s)&= V_{cb}V_{cs}^*(T_{c\bar cs}+P^c_s-P^t_s)
+V_{ub}V^*_{us}(P^u_s-P^t_s), \cr
A(u\bar u s)&= V_{cb}V_{cs}^*(P^c_s-P^t_s)
+V_{ub}V^*_{us}(T_{u\bar us}+P^u_s-P^t_s), \cr
A(s\bar s s)&= V_{cb}V_{cs}^*(P^c_s-P^t_s)+V_{ub}V^*_{us}(P^u_s- P^t_s).
\cr}}
In these expressions only differences of penguin contributions occur,
which makes the cancellation of the ultraviolet divergences of these
diagrams explicit. Furthermore, the second term has a CKM coefficient that
is much smaller, by $\O(\lambda^2)$, than the first. Hence this grouping 
is useful in classifying the expected CP violation in decay.  (Note that terms
$b\ra d \bar d s$, which have only penguin contributions, mix strongly
with the $u\bar u s$ terms and hence cannot be separated from them. Thus
$P$ terms in  $A(u \bar u s)$ include
contributions from both $d \bar d s$ and  $u\bar u s$ diagrams.)
 
In the case of $q \bar q d $ decays the three CKM coefficients are 
of similar magnitude. The convention is then to retain the
$V_{tb}V^*_{td}$ term because, in the Standard Model, the phase
difference between this weak phase and half the mixing weak phase is
zero. Thus only one unknown weak phase enters the calculation of the
interference between decays with and without mixing.
We can choose to eliminate which of the other terms does not
have a tree contribution. In the cases $q=s$ or $d$, since neither has a
tree contribution either term can be removed. Thus we write
\eqn\qqdtype{\eqalign{
A(c\bar c d)&=V_{tb}V_{td}^*(P^t_d-P^u_d)
+ V_{cb}V_{cd}^*(T_{c\bar c d}+P^c_d-P^u_d), \cr
A(u\bar u d)&=V_{tb}V_{td}^*(P^t_d-P^c_d)
 +V_{ub}V_{ud}^*(T_{u\bar u d}+P^u_d - P^c_d), \cr
A(s\bar s d)&=V_{tb}V_{td}^*(P^t_d-P^u_d)+V_{cb}V_{cd}^*(P^c_d-P^u_d).
\cr}}
Again only differences of penguin amplitudes occur. Furthermore the
difference of penguin terms that occurs in the second term would vanish 
if the charm and up quark masses were equal, and thus is GIM suppressed
\ref\GIM{S.L. Glashow, J. Iliopoulos and L. Maiani,
 Phys. Rev. D2 (1970) 1285.}. 
However, even in modes with no tree contribution, $(s\bar s d)$, 
the interference of the two terms can still give significant 
CP violation in the interference of decays with and without mixing.
 
The penguin processes all involve the emission of a neutral boson,
either a gluon (strong penguins) or a photon or $Z$ boson (electroweak
penguins). Excluding the CKM coefficients, the ratio of the contribution
from the difference between a top and light quark strong penguin diagram
to the contribution from a tree diagram is of order
\eqn\pengtree{
r_{PT} = { P^t-P^{light}\over T_{q \bar q q^\prime
 }}\approx{\alpha_s\over12\pi}\ln{m_t^2\over m_b^2}.}
This is a factor of $\O(0.03)$. However this estimate does not include
the effect of hadronic matrix elements, which are the probability
factor to produce a particular final state particle content from a
particular quark content. Since this probability differs for different
kinematics, color flow and spin structures, it can be different for tree
and penguin contributions and may partially compensate the coupling
constant suppression of the penguin term. Recent CLEO results on 
$BR(B\ra K\pi)$ and $BR(B\ra\pi\pi)$
\ref\CLEOpen{R. Godang {\it et al.}, CLEO Collaboration, 
 Phys. Rev. Lett. 80 (1998) 3456, hep-ex/9711010.}\ 
suggest that the matrix element of penguin operators is indeed enhanced
compared to that of tree operators. The enhancement could be by a factor
of a few, leading to
\eqn\pentre{r_{PT}\sim\lambda^2-\lambda.}
(Note that $r_{PT}$ does not depend on the CKM parameters.
We use powers of the Wolfenstein parameter $\lambda$ to 
quantify our estimate for $r_{PT}$ is order to simplify
the comparison between the size of CP violation in decay
and CP violation in the interference between decays with
and without mixing.) Electroweak penguin difference
terms are even more suppressed since they have an $\alpha_{\rm EM}$ or 
$\alpha_W$ instead of the $\alpha_s$ factor in \pengtree, but certain 
$Z$-contributions are enhanced by the large top
quark mass and so can be non-negligible.
 
We thus classify $B$ decays into four classes. Classes (i) and (ii) are
expected to have relatively small CP violation in decay and hence are
particularly interesting for extracting CKM parameters from
interference of decays with and without mixing. In the remaining
two classes, CP violation in decay could be significant and the neutral 
decay asymmetries cannot be cleanly interpreted in terms of CKM phases.

\item{(i)} Decays dominated by a single term: $b\ra c\bar cs$ and $b\ra
s\bar ss$. The Standard Model cleanly predicts very small
CP violation in decay: $\O(\lambda^4-\lambda^3)$ for $b\ra c\bar cs$
and $\O(\lambda^2)$ for $b\ra s\bar ss$.
Any observation of large CP asymmetries in charged $B$ decays
for these channels would be a clue to physics beyond the Standard Model. 
The corresponding neutral modes have cleanly predicted relationships
between CKM parameters and the measured asymmetry from interference
between decays with and without mixing. The modes
$B\ra\psi K$ and $B\ra\phi K$ are examples of this class.
 
\item{(ii)} Decays with a small second term: $b\ra c\bar cd$ and $b\ra
u \bar ud$. The expectation that penguin-only contributions are supressed
compared to tree contributions suggests that these modes will have small
effects of CP violation in decay, of $\O(\lambda^2-\lambda)$, and an 
approximate prediction for the relationship between measured asymmetries 
in neutral decays and CKM
phases can be made. Examples here are $B\ra\ DD$ and $B\ra\pi\pi$.
 
\item{(iii)} Decays with a suppressed tree contribution: $b\ra u \bar us$.
The tree amplitude is suppressed by small mixing angles, $V_{ub}V_{us}$.
The no-tree term may be comparable or even dominate and give large
interference effects. An example is $B\ra\rho K$.
 
\item{(iv)} Decays with no tree contribution: $b\ra s\bar s d$. Here the
interference comes from penguin contributions with different charge 2/3
quarks in the loop and gives CP violation in decay that could be as large 
as 10\%
\nref\Flei{R. Fleischer, Phys. Lett. B341 (1995) 205.}%
\nref\BuFl{A.J. Buras and R. Fleischer, Phys. Lett. B341 (1995) 379.}%
\refs{\Flei,\BuFl}. An example is $B\ra KK$.

Note that if the penguin enhancement is significant,
then some of the decay modes listed in class (ii) might actually fit 
better in class (iii). For example, it is possible that $b\ra u\bar ud$ 
decays have comparable contributions from tree and penguin amplitudes.
On the other hand, this would also mean that some modes listed in class 
(iii) could be dominated by a single penguin term. For such cases
an approximate relationship between measured
asymmetries in neutral decays and CKM phases can be made. 

\subsec{CP violation in the interference between $B$ decays with and 
without mixing}
 Let us first discuss an example of class (i), $B\ra\psi K_S$. A new
ingredient in the analysis is the effect of $K-\bar K$ mixing. For decays
with a single $K_S$ in the final state, $K-\bar K$ mixing is essential
because $B^0\ra K^0$ and $\bar B^0\ra\bar K^0$, and interference is
possible only due to $K-\bar K$ mixing. This adds a factor of
\eqn\qpK{
\left({p\over q}\right)_K=
{V_{cs}V_{cd}^*\over V_{cs}^*V_{cd}}\omega_K^*}
into $(\bar A/A)$. The quark subprocess in $\bar B^0\ra\psi\bar K^0$ is
$b\ra c\bar cs$ which is dominated by the $W$-mediated tree diagram:
\eqn\ApsiK{
{\bar A_{\psi K_S}\over A_{\psi K_S}}= \eta_{\psi K_S}
\left({V_{cb}V_{cs}^*\over V_{cb}^*V_{cs}}\right)
\left({V_{cs}V_{cd}^*\over V_{cs}^*V_{cd}}\right)\omega_B^*.}
The CP-eigenvalue of the state is $\eta_{\psi K_S} = -1$.
Combining \qpSM\ and \ApsiK, we find
\eqn\lampsiK{
\lambda(B\ra\psi K_S)=-\left({V_{tb}^*V_{td}\over
V_{tb}V_{td}^*}\right)\left({V_{cb}V_{cs}^*\over V_{cb}^*V_{cs}}\right)
\left({V_{cd}^*V_{cb}\over V_{cd}V_{cb}^*}\right)
\ \Longrightarrow\ \Im\lambda_{\psi K_S}=\sin(2\beta).}
 
The second term in \qqstype\ is of order $\lambda^2r_{PT}$
for this decay and thus eq.  \lampsiK\ is clean of hadronic
uncertainties to $\O(10^{-3})$. Consequently, this measurement can give 
the theoretically cleanest determination of a CKM parameter, even 
cleaner than the determination of $|V_{us}|$ from $K\ra\pi\ell\nu$. (If
BR($K_L\ra\pi\nu\bar\nu$) is measured, it will give a comparably clean
determination of $\eta$.)
 
A second example of a theoretically clean mode in class (i) is
$B\ra\phi K_S$. The quark subprocess involves FCNC
and cannot proceed via a tree level SM diagram. The leading
contribution comes from penguin diagrams. The two terms in eq.
\qqstype\ are now both differences of penguins but the second term
is CKM suppressed and thus of $\O(\lambda^2)$ compared to the first. Thus
CP violation in the decay is at most a few percent, and can be
neglected in the analysis of asymmetries in this channel.  The analysis
is similar to the $\psi K_S$ case, and the asymmetry is proportional to
$\sin(2\beta)$:
 
The same quark subprocesses give theoretically clean CP
asymmetries also in $B_s$ decays. These asymmetries are, however,
very small since the relative phase between the mixing amplitude
and the decay amplitudes ($\beta_s$ defined below) is very small.
 
The best known example of class (ii) is $B\ra\pi\pi$. The quark subprocess
is $b\ra u\bar ud$ which is dominated by the $W$-mediated tree diagram.
Neglecting for the moment the second, pure penguin, term in eq.
\qqdtype\ we find
\eqn\Apipi{
{\bar A_{\pi\pi}\over A_{\pi\pi}}=\eta_{\pi\pi}{V_{ub}V_{ud}^*\over
 V_{ub}^*V_{ud}}\omega_B^*.}
The CP eigenvalue for two pions is $+1$.
Combining \qpSM\ and \Apipi, we get
\eqn\lampipi{
\lambda(B\ra\pi^+\pi^-)=\left({V_{tb}^*V_{td}\over V_{tb}V_{td}^*}\right)
\left({V_{ud}^*V_{ub}\over V_{ud}V_{ub}^*}\right)
\ \Longrightarrow\ \Im\lambda_{\pi\pi}=\sin(2\alpha).}
The pure penguin term in eq. \qqdtype\ has a weak phase,
$\arg(V_{td}^*V_{tb})$, different from the term with the tree
contribution, so it modifies both $\Im\lambda$ and (if there are
non-trivial strong phases) $|\lambda|$. The recent CLEO results 
mentioned above suggest that the penguin contribution
to $B\ra\pi\pi$ channel is significant, probably  $10\%$ or more. This
then introduces CP violation in decay, unless the strong phases
cancel (or are zero, as suggested by factorization arguments).
The resulting hadronic uncertainty can be eliminated using isospin analysis
\ref\GrLo{M. Gronau and D. London, Phys. Rev. Lett. 65 (1990) 3381.}.
This requires a measurement of the rates for the
isospin-related channels $B^+ \ra \pi^+ \pi^0$ and $B^0 \ra \pi^0\pi^0$
as well as the corresponding CP-conjugate processes. The rate for
$\pi^0\pi^0$ is expected to be small and the measurement is difficult,
but even an upper bound on this rate can be used to limit the magnitude
of hadronic uncertainties
\ref\GrQu{Y. Grossman and H. Quinn,
 Phys. Rev. D58 (1998) 017504, hep-ph/9712306.}.

Related but slightly more complicated channels with the same underlying
quark structure are $B\ra\rho^0\pi^0$ and $B\ra a_1^0\pi^0$. Again an
analysis involving the isospin-related channels can be used to help
eliminate hadronic uncertainties from CP violations in the decays
\nref\LNQS{
H.J. Lipkin, Y. Nir, H.R. Quinn and A. Snyder, Phys. Rev. D44 (1991) 1454.}%
\nref\SnQu{H.R. Quinn and A. Snyder, Phys. Rev. D48 (1993) 2139.}%
\refs{\LNQS,\SnQu}.
Channels such as $\rho\rho$ and $a_1\rho$ could in principle also be
studied, using angular analysis to determine the mixture of
CP-even and CP-odd contributions.
  
The analysis of $B\ra D^+D^-$ proceeds along very similar lines.
The quark subprocess here is $b\ra c\bar cd$, and so the tree
contribution gives
\eqn\lamDD{
\lambda(B\ra D^+D^-)=
\eta_{D^+D^-}\left({V_{tb}^*V_{td}\over V_{tb}V_{td}^*}\right)
\left({V_{cd}^*V_{cb}\over V_{cd}V_{cb}^*}\right)
\ \Longrightarrow\ \Im\lambda_{DD}=-\sin(2\beta).}
since $\eta_{D^+D^-}=+1$.
Again, there are hadronic uncertainties due to the pure penguin term in
\qqdtype, but they are estimated to be small. 
 
 
In all cases the above discussions have neglected the distinction
between strong penguins and electroweak penguins. The CKM phase
structure of both types of penguins is the same. The only place where
this distinction becomes important is when an isospin argument is used
to remove hadronic uncertainties due to penguin contributions. These
arguments are based on the fact that gluons have isospin zero, and
hence strong penguin processes have definite $\Delta I$. Photons and
$Z$-bosons on the other hand contribute to more than one $\Delta I$
transition and hence cannot be separated from tree terms by isospin
analysis. In most cases electroweak penguins are small, typically no more
than ten percent of the corresponding strong penguins and so their
effects can safely be neglected. However in cases (iii) and (iv), where tree
contributions are small or absent, their effects may need to be considered.
(A full review of the role of electroweak penguins in $B$
decays has been given in ref. 
\ref\ewpenreview{R. Fleischer,
 Int. J. Mod. Phys. A12 (1997) 2459, hep-ph/9612446.}.) 

\subsec{Unitarity Triangles}
One can obtain an intuitive understanding of the Standard Model
CP violation in the interference between decays with and without mixing
by examining the unitarity triangles. It is instructive to draw the 
three triangles, \Unitds, \Unitsb\ and \Unitdb, knowing the experimental
values (within errors) for the various $|V_{ij}|$. In the first
triangle \Unitds, one side is of $\O(\lambda^5)$ and therefore 
much shorter than the other, $\O(\lambda)$, sides. In the second 
triangle \Unitsb, one side is of $\O(\lambda^4)$ and therefore shorter 
than the other, $\O(\lambda^2)$, sides. In the third triangle \Unitdb, 
all sides have lengths of $\O(\lambda^3)$. The first two triangles then 
almost collapse to a line while the third one is open. 

Let us examine the CP asymmetries in the leading decays into
final CP eigenstates. For the $B$ mesons, the size of these 
asymmetries ({\it e.g.} $\Im\lambda_{\psi K_S}$) depends on 
$\beta$ because it gives the difference between half the phase of 
the $B-\bar B$ mixing amplitude and the phase of the decay
amplitudes. The form of the third unitarity triangle, \Unitdb, 
implies that $\beta=\O(1)$, which explains 
why these asymmetries are expected to be large. 

It is useful to define the analog phases for the
$B_s$ meson, $\beta_s$, and the $K$ meson, $\beta_K$:
\eqn\bbangles{
\beta_s\equiv\arg\left[-{V_{ts}V_{tb}^*\over V_{cs}V_{cb}^*}\right],\ \ \
\beta_K\equiv\arg\left[-{V_{cs}V_{cd}^*\over V_{us}V_{ud}^*}\right].}
The angles $\beta_s$ and $\beta_K$ can be seen to be the small angles of
the second and first unitarity triangles, \Unitsb\ and \Unitds, respectively. 
This gives an intuitive understanding of why CP violation is small in the 
leading $K$ decays (that is $\epsK$ measured in $K\rightarrow\pi\pi$
decays) and is expected to be small in the leading $B_s$ decays ({\it e.g.}
$B_s\rightarrow\psi\phi$). Decays related to the short sides of these 
triangles are rare but could exhibit
significant CP violation. Actually, the large angles in the \Unitds\
triangle are approximately $\beta$ and $\pi-\beta$, which explains why
CP violation in $K\ra\pi\nu\bar\nu$ is related to $\beta$ and expected
to be large. The large angles in the \Unitsb\
triangle are approximately $\gamma$ and $\pi-\gamma$. This explains
why the CP asymmetry in $B_s\ra\rho K_S$ is related to $\gamma$
and expected to be large. (Note, however, that this mode gets
comparable contributions from penguin and tree diagrams and does
not give a clean CKM measurement \BuFl.) 

\newsec{CP Violation Beyond the Standard Model}
The Standard Model picture of CP violation is rather unique and highly
predictive. In particular, we would like to point out the following features:
\item{(i)} CP is broken explicitly.
\item{(ii)} All CP violation arises from a single phase, that is $\dKM$.
\item{(iii)} The measured value of $\epsK$ requires that $\dKM$
is of order one. (In other words, CP is not an approximate symmetry of
the Standard Model.)
\item{(iv)} The values of all other CP violating observables
can be predicted. In particular, CP violation in $B\ra\psi K_S$
(and similarly various other CP asymmetries in $B$ decays), 
and in $K\ra\pi\nu\bar\nu$ are expected to be of order one.

The commonly repeated statement that CP violation is one of the least
tested aspects of the Standard Model is well demonstrated by the fact
that none of the above features necessarily holds in the presence
of New Physics. In particular, there are viable models of new physics
({\it e.g.} certain supersymmetric models) with the following features:
\item{(i)} CP is broken spontaneously.
\item{(ii)} There are many CP violating phases (even in the low
energy effective theory).
\item{(iii)}  CP is an approximate symmetry, with all CP violating
phases small (usually $10^{-3}\lsim\phi_{\rm CP}\lsim10^{-2}$).
\item{(iv)} Values of CP violating observables
can be predicted and could be very different from the Standard
Model predictions (except, of course, $\epsK$). In particular, 
$\Im\lambda_{\psi K_S}$ and $a_{\pi\nu\bar\nu}$ could both be $\ll1$.     

To understand how the Standard Model predictions could be modified 
by New Physics, we will focus on CP violation in the interference
between decays with and without mixing. As explained above, it is
this type of CP violation which, due to its theoretical cleanliness,
may give unambiguous evidence for New Physics most easily.

\subsec{CP Violation as a Probe of Flavor Beyond the Standard Model}
Let us consider five specific CP violating observables.

(i) $\Im\lambda_{\psi K_S}$, the CP asymmetry in $B\rightarrow\psi K_S$.
This measurement will cleanly determine the relative phase between
the $B-\bar B$ mixing amplitude and the $b\ra c\bar cs$ decay amplitude
($\sin2\beta$ in the Standard Model).
The $b\ra c\bar cs$ decay has Standard Model tree contributions and
therefore is very unlikely to be significantly affected by new physics.
On the other hand, the mixing amplitude can be easily modified
by new physics. We parametrize such a modification by a phase $\theta_d$:
\eqn\apksNP{\Im\lambda_{\psi K_S}=\sin[2(\beta+\theta_d)].}

(ii) $\Im\lambda_{\phi K_S}$, the CP asymmetry in $B\rightarrow\phi K_S$.
This measurement will cleanly determine the relative phase between
the $B-\bar B$ mixing amplitude and the $b\ra s\bar ss$ decay amplitude.
The $b\ra s\bar ss$ decay has only Standard Model penguin contributions
and therefore is sensitive to new physics. We parametrize the modification 
of the decay amplitude by a phase $\theta_A$
\ref\GrWo{Y. Grossman and M.P. Worah, Phys. Lett. B395 (1997) 241.}:
\eqn\aphksNP{\Im\lambda_{\phi K_S}=\sin[2(\beta+\theta_d+\theta_A)].}

(iii) $a_{\pi\nu\bar\nu}$, the CP violating ratio of $K\rightarrow
\pi\nu\bar\nu$ decays. This measurement will cleanly determine the 
relative phase between the $K-\bar K$ mixing amplitude and the 
$s\ra d\nu\bar\nu$ decay amplitude.
The experimentally measured small value of $\epsK$ requires that
the phase of the $K-\bar K$ mixing amplitude is not modified from
the Standard Model prediction. On the other hand, the decay,
which in the Standard Model is a loop process with small mixing angles, 
can be easily modified by new physics. 

(iv) $\Im(\lambda_{K^-\pi^+})$, the CP violating
quantity in $D\rightarrow K^-\pi^+$ decay. The ratio
\eqn\defaD{
a_{D\ra K\pi}={\Im(\lambda_{K^-\pi^+})\over|\lambda_{K^-\pi^+}|}}
depends on the relative phase between the $D-\bar D$ mixing amplitude 
and the $c\ra d\bar su$ decay amplitude. Within the Standard Model, 
this decay channel is tree level. It is unlikely that it is 
affected by new physics. On the other hand, the mixing amplitude can 
be easily modified by new physics.

(v) $d_N$, the electric dipole moment of the neutron.
We did not discuss this quantity so far because, unlike
CP violation in meson decays, flavor changing couplings
are not necessary for $d_N$. In other words, the CP violation
that induces $d_N$ is {\it flavor diagonal}. It does in general
get contributions from flavor changing physics, but it could
be induced by sectors that are flavor blind. Within the
Standard Model (and ignoring the strong CP angle $\theta_{\rm QCD}$),
the contribution from $\delta_{\rm KM}$ arises at the three loop
level and is at least six orders of magnitude below the
experimental bound \PDG\ $d_N^{\exp}$,
\eqn\dnexp{d_N^{\rm exp}=1.1\times10^{-25}\ e\ {\rm cm}.}

The various CP violating observables discussed above are sensitive
then to new physics in the mixing amplitudes for the $B-\bar B$ 
and $D-\bar D$ systems, in the decay amplitudes for $b\rightarrow s\bar ss$
and $s\ra d\nu\bar\nu$ channels and to flavor diagonal CP violation.
If information about all these processes becomes available
and deviations from the Standard Model predictions are found,
we can ask rather detailed questions about the nature of
the new physics that is responsible to these deviations:
\item{(i)} Is the new physics related to the down sector?
the up sector? both?
\item{(ii)} Is the new physics related to $\Delta B=1$ processes? 
$\Delta B=2$? both?
\item{(iii)} Is the new physics related to the third generation?
to all generations?
\item{(iv)} Are the new sources of CP violation flavor changing?
flavor diagonal? both?

It is no wonder then that with such rich information,
flavor and CP violation provide an excellent probe of new physics.

\subsec{Supersymmetry}
 
A generic supersymmetric extension of the Standard Model contains a host
of new flavor and CP violating parameters. (For reviews on supersymmetry see
\nref\nilles{H.P. Nilles, Phys. Rep. 110 (1984) 1.}%
\nref\haberkane{H.E. Haber and G.L. Kane, Phys. Rep. 117 (1985) 75.}%
\nref\barbieri{R. Barbieri, Riv. Nuovo Cim. 11 (1988) 1.}%
\nref\HaberTASI{H.E. Haber, SCIPP 92/33, Lectures given at TASI 92.}%
refs. \refs{\nilles-\HaberTASI}. The following section is based on
\ref\GNR{Y. Grossman, Y. Nir and R. Rattazzi, hep-ph/9701231, to appear
in {\it Heavy Flavours II}, eds. A.J. Buras and M. Lindner (World
Scientific).}.) The requirement of consistency
with experimental data provides strong constraints on many of these
parameters. For this reason, the physics of flavor and CP violation
has had a profound impact on supersymmetric model building. A discussion
of CP violation in this context can hardly avoid addressing the flavor
problem itself.  Indeed, many of the supersymmetric models that we
analyze below were originally aimed at solving flavor problems. 

As concerns CP violation, one can distinguish two classes of experimental
constraints. First, bounds on nuclear and atomic electric dipole moments
determine what is usually called the {\it supersymmetric CP problem}.
Second, the physics of neutral mesons and, most importantly, the small
experimental value of $\epsK$ pose the {\it supersymmetric $\epsK$
problem}. The latter is closely related to the flavor structure of 
supersymmetry.

The contribution to the CP violating $\epsK$ parameter in the neutral $K$
system is dominated by diagrams involving $Q$ and $\bar d$ squarks in the
same loop
\nref\DNW{J. Donoghue, H. Nilles and D. Wyler,
 Phys. Lett. B128 (1983) 55.}%
\nref\GaMa{F. Gabbiani and A. Masiero, Nucl. Phys. B322 (1989) 235.}%
\nref\HKT{J.S. Hagelin, S. Kelley and T. Tanaka,
 Nucl. Phys. B415 (1994) 293.}%
\nref\GMS{E. Gabrielli, A. Masiero and L. Silvestrini,
 Phys. Lett. B374 (1996) 80.}%
\nref\GGMS{F. Gabbiani, E. Gabrielli, A. Masiero and L. Silvestrini,
Nucl. Phys. B477 (1996) 321.}%
\refs{\DNW-\GGMS}. The corresponding effective four-fermi operator
involves fermions of both chiralities, so that its matrix elements
are enhanced by $\O(m_K/m_s)^2$ compared to the chirality conserving
operators. For $m_{\tilde g}\simeq m_Q \simeq m_D= \tilde m$ (our results
depend only weakly on this assumption) and focusing on the contribution
from the first two squark families, one gets (we use the results in
ref. \GGMS)
\eqn\epsKScon{{(\Delta m_K\epsK)^{\rm SUSY}\over
\Delta m_K\epsK}\sim10^7\ 
\left ({300 \ \gev\over\tilde m}\right)^2\left|
{(\delta m_D^2)_{12} \over\tilde m^2}\right|^2 |K_{12}^d|^2\sin\phi,}
where $\tilde m$ is the typical scale of squark and gluino masses,
$(\delta m_D^2)_{12}$ is the mass-squared difference between the
first two down squark generations, $K_{12}^d$ is the mixing angle
in the gluino-quark-squark coupling and $\phi$ is the relevant
CP violating phase in the mixing.  
In a generic supersymmetric framework, we expect $\tilde m=\O(m_Z)$,
$\delta m_{D}^2/\tilde m^2=\O(1)$, $K_{ij}^d=\O(1)$ and
$\sin\phi=\O(1)$. Then the constraint \epsKScon\ is generically violated
by about seven orders of magnitude. Eq. \epsKScon\ also shows what are 
the possible solutions to the supersymmetric flavor and CP problems:

(i){\it Universality}: At some high scale, the soft supersymmetry
breaking terms are universal. In other words, the different
squark generations are degenerate
\nref\DiGe{S. Dimopoulos and H. Georgi, Nucl. Phys. B193 (1981) 150.}%
\nref\Saka{N. Sakai, Z. Phys. C11 (1981) 153.}%
\refs{\DiGe-\Saka}. There are two very different ways to achieve 
such a situation. First, the mechanism that communicates supersymmetry 
breaking to the observable sector could be flavor blind. This is the 
case with gauge mediated supersymmetry breaking
\nref\DiNe{M. Dine and A. Nelson, Phys. Rev. D48 (1993) 1277.}%
\nref\DNeS{M. Dine and A. Nelson and Y. Shirman,
 Phys. Rev. D51 (1994) 1362.}%
\nref\DNNS{M. Dine, A. Nelson, Y. Nir and Y. Shirman,
 Phys. Rev. D53 (1996) 2658.}%
 \nref\DNiS{M. Dine, Y. Nir and Y. Shirman,
 Phys. Rev. D55 (1997) 1501.}%
\refs{\DiNe-\DNiS}, but it is also possible (though not generic) 
that similar boundary conditions occur when supersymmetry breaking 
is communicated to the observable sector up at the Planck scale
\nref\CAN{A.H. Chamseddine, R. Arnowitt and P. Nath,
 Phys. Rev. Lett. 49 (1982) 970; Nucl. Phys. B227 (1983) 1219.}%
\nref\BFS{R. Barbieri, S. Ferrara and C.A. Savoy,
 Phys. Lett. B119 (1982) 343.}%
\nref\HLW{L. Hall, J. Lykken and S. Weinberg,
 Phys. Rev. D27 (1983) 235.}%
\nref\EKN{J. Ellis, C. Kounnas and D.V. Nanopoulos, Nucl. Phys. B247
 (1984) 373.}%
\nref\LaRo{M. Lanzagorta and G.G. Ross, Phys. Lett. B364 (1995) 163.}%
\nref\KaLo{V. Kaplunovsky and J. Louis, Phys. Lett. B306 (1993) 269.}%
\nref\BML{R. Barbieri, J. Louis and M. Moretti,
 Phys. Lett. B312 (1993) 451; (E) {\it ibid.} B316 (1993) 632.}%
\refs{\CAN-\BML}. RGE effects will introduce some splitting
at low energy which, for the first two squark generations, is
typically of $\O(m_c^2/m_W^2)$. Second, the Yukawa hierarchy could
be a result of a non-Abelian flavor symmetry with the first two 
generations forming a doublet
\nref\DKL{M. Dine, A. Kagan and R.G. Leigh, Phys. Rev. D48 (1993) 4269.}%
\nref\PoSe{P. Pouliot and N. Seiberg, Phys. Lett. B318 (1993) 169.}%
\nref\HaMu{L.J. Hall and H. Murayama, Phys. Rev. Lett. 75 (1995) 3985.}%
\nref\PoTo{A. Pomarol and D. Tommasini,
 Nucl. Phys. B466 (1996) 3, hep-ph/9507462.}%
\nref\BDH{R. Barbieri, G. Dvali and L.J. Hall,
 Phys. Lett. B377 (1996) 76, hep-ph/9512388.}%
\nref\CHM{C. Carone, L.J. Hall and H. Murayama,
 Phys. Rev. D54 (1996) 2328, hep-ph/9602364.}%
\nref\Zurab{Z.G. Berezhiani, hep-ph/9609342.}%
\nref\Raby{R. Barbieri, L.J. Hall, S. Raby and A. Romanino,
 Nucl. Phys. B493 (1997) 3, hep-ph/9610449.}%
\nref\Gali{G. Eyal, hep-ph/9807308.}%
\refs{\DKL-\Gali}. In this framework, the first two squark generations 
are approximately degenerate with splitting which could be as high as
$\O(\lambda^2)$. The third generation could be widely
split from the first two.

(ii) {\it Alignment}
\nref\NiSe{Y. Nir and N. Seiberg,
 Phys. Lett. B309 (1993) 337, hep-ph/9304307.}%
\nref\LNSb{M. Leurer, Y. Nir and N. Seiberg,
 Nucl. Phys. B420 (1994) 468, hep-ph/9410320.}%
\nref\RaNi{Y. Nir and R. Rattazzi,
 Phys. Lett. B382 (1996) 363, hep-ph/9603233.}%
\refs{\NiSe-\RaNi}: The mixing angles in the gluino-quark-squark
couplings are small. This is usually achieved in models where the
Yukawa hierarchy is explained by Abelian flavor symmetries. 
In the symmetry limit, both the quark mass matrices
and the squark mass-squared matrices are diagonal, so that mixing
is suppressed by small breaking parameters. Typically, the alignment
is required to be very precise between the first two down generations,
while all other supersymmetric mixing angles are similar to the 
corresponding CKM angles.

(iii) {\it Heavy Squarks}
\nref\DKS{M. Dine, A. Kagan and S. Samuel, Phys. Lett. B243 (1990) 250.}%
\nref\CKN{A.G. Cohen, D.B. Kaplan and A.E. Nelson,
 Phys. Lett. B388 (1996) 588, hep-ph/9607394.}%
\nref\CKLN{A.G. Cohen, D.B. Kaplan, F. Lepeintre and A.E. Nelson,
 Phys. Rev. Lett. 78 (1997) 2300, hep-ph/9610252.}%
\refs{\DKL,\PoTo,\DKS-\CKLN}:
If the masses of the first and second generation
squarks $m_i$ are larger than the other soft masses, $m_i^2\sim 100\,
\tilde m^2$, then the Supersymmetric CP problem is solved and the $\epsK$
problem is relaxed (but not eliminated). This does not necessarily lead 
to naturalness problems, since these two generations are
almost decoupled from the Higgs sector.

(iv) {\it Approximate CP}
\nref\BaBa{K.S. Babu and S.M. Barr, Phys. Rev. Lett. 72 (1994) 2831.}%
\nref\abefre{S.A. Abel and J.M. Frere, Phys. Rev. D55 (1997) 1623.}%
\nref\Eyal{G. Eyal and Y. Nir, Nucl. Phys. B528 (1998) 21, hep-ph/9801411.}%
\refs{\BaBa-\Eyal}.:
Both supersymmetric CP problems are solved if CP is an approximate
symmetry, broken by a small parameter of order $10^{-3}$.
Of course, some mechanism to solve the supersymmetric flavor problems
has to be invoked.

Measurements of CP violation will provide us with an
excellent probe of the flavor and CP structure of supersymmetry.
This is clearly demonstrated in the Table below. 
\vskip 1cm
 
\begintable
Model & $d_N/d_N^{\rm exp}$ & $\theta_d$ & $\theta_A$ &
$a_{D\ra K\pi}$ & $a_{K\ra\pi\nu\bar\nu}$ \crthick
Standard Model & $\lsim10^{-6}$ & $0$ & $0$ & $0$ & $\O(1)$ \nr
Exact Universality & $\lsim10^{-6}$ & $0$ & $0$ & $0$ & =SM \nr
Approximate CP & $\sim10^{-1}$ & $-\beta$ & $0$ & $\O(10^{-3})$
& $\O(10^{-5})$ \nr
Alignment & $\gsim10^{-3}$ & $\O(0.2)$ & $\O(1)$ &
$\O(1)$ & $\approx$SM \nr
Approx. Universality & $\gsim10^{-2}$ & $\O(0.2)$ & $\O(1)$ &
$0$ & $\approx$SM \nr
Heavy Squarks & $\sim10^{-1}$ & $\O(1)$ & $\O(1)$ &
$\O(10^{-2})$ & $\approx$SM \endtable

\centerline{CP violating observables in various classes
of Supersymmetric flavor models.} 

\subsec{Final Comments}
The unique features of CP violation are well demonstrated by
examining the CP asymmetry in $B\rightarrow\psi K_S$, 
$\Im\lambda_{\psi K_S}$, and CP violation in $K\rightarrow\pi
\nu\bar\nu$, $\Im\lambda_{\pi\nu\bar\nu}$. Model independently,
$\Im\lambda_{\psi K_S}$ measures the relative phase between the
$B-\bar B$ mixing amplitude and the $b\rightarrow c\bar cd$ decay
amplitude (more precisely, the $b\rightarrow c\bar cs$ decay
amplitude times the $K-\bar K$ mixing amplitude), while
$\Im\lambda_{\pi\nu\bar\nu}$ measures the relative phase between the
$K-\bar K$ mixing amplitude and the $s\rightarrow d\nu\bar\nu$ decay
amplitude. We would like to emphasize the following three points:

(i) {\it The two measurements are theoretically clean to better
than} $\O(10^{-2})$. Thus they can provide the most accurate
determination of CKM parameters. 

(ii) {\it As concerns CP violation, the Standard Model is a uniquely
predictive model}. In particular, it predicts that the seemingly
unrelated $\Im\lambda_{\psi K_S}$ and $\Im\lambda_{\pi\nu\bar\nu}$
measure the same parameter, that is the angle $\beta$ of the
unitarity triangle.

(iii) {\it In the presence of New Physics, there is in general no
reason for a relation between $\Im\lambda_{\psi K_S}$ and 
$\Im\lambda_{\pi\nu\bar\nu}$}. Therefore, a measurement of both
will provide a sensitive probe of New Physics.

\bigskip
\noindent
{\bf Acknowledgements}
\smallskip
\noindent
My understanding of flavor physics and CP violation has benefitted
from numerous discussions with Helen Quinn. Large parts of this
review are based on previous reviews written in collaboration
with her. Parts of this review are based on contributions
to the BaBar Physics Book that were written in collaboration
with Gerald Eigen, Ben Grinstein, Stephane Plaszczynski and
Marie-Helene Schune. Other parts of this review are based
on a review written in collaboration with Yuval Grossman and
Riccardo Rattazzi. I thank Sven Bergmann, Galit Eyal, Eilam Gross
and Yael Shadmi for their help in preparing these lectures.
Y.N. is supported in part by the United States $-$ Israel Binational 
Science Foundation (BSF), by the Israel Science
Foundation and by the Minerva Foundation (Munich).

\listrefs
\end